\documentclass[11pt]{article}

\usepackage{amsmath,amsfonts,amssymb,amsthm,bbm,stmaryrd}
\usepackage[dvips]{graphicx}
\usepackage{latexsym}
\usepackage{psfrag}
\usepackage[latin1]{inputenc}
\usepackage{hyperref}
\usepackage{color}
\usepackage{enumitem}
\setlistdepth{9}
\numberwithin{equation}{section}
\usepackage{float}
\usepackage{multirow}

\newcommand{\be}{\begin{equation}}
\newcommand{\ee}{\end{equation}}
\newcommand{\barray}{\begin{array}}
\newcommand{\earray}{\end{array}}
\newcommand{\bea}{\begin{eqnarray}}
\newcommand{\eea}{\end{eqnarray}}
\newcommand{\bs}{\begin{subequations}}
\newcommand{\es}{\end{subequations}}

\newcommand{\bit}{\begin{itemize}}
\newcommand{\eit}{\end{itemize}}
\newcommand{\bd}{\begin{description}}
\newcommand{\ed}{\end{description}}

\def\nn{\nonumber}

\newcommand{\re}{\mathrm{Re}}

\def\la{\langle}
\def\ra{\rangle}

\newcommand{\mat} [4] {\left ( \begin{array}{cc}{#1}&{#2}\\{#3}&{#4} \end{array} \right ) }

\def\w{\wedge}
\newcommand{\p}{\partial}
\newcommand{\na}{\nabla}

\newcommand{\R}{\mathbb{R}}
\newcommand{\C}{\mathbb{C}}

\newcommand{\f}{\frac}
\newcommand{\tl}{\tilde}

\renewcommand{\a}{\alpha} \renewcommand{\b}{\beta} \newcommand{\g}{\gamma}  
\renewcommand{\d}{\delta}  \newcommand{\eps}{\epsilon} 
 \renewcommand{\th}{\theta}  \newcommand{\vth}{\vartheta} 
  \renewcommand{\k}{\kappa}  \renewcommand{\l}{\lambda}
\let\m=\mu    \let\n=\nu   \let\r=\rho \let\om=\omega
 \newcommand{\s}{\sigma}  \renewcommand{\t}{\tau}    
\let\G=\Gamma \let\D=\Delta    
\let\Si=\Sigma

\newcommand{\SL}{\mathrm{SL}(2,\mathbb{C})}

\newcommand{\bigG}{{\Diff_l(\cN)}}

\def\cN{{\cal N}}

\newcommand{\scri}{{\cal I}}
\newcommand{\os}[1]{\overset{\circ}{#1}}

\usepackage{slashed}
\newcommand{\sd}{\slashed{\delta}}

\newcommand{\eqons}{\,\hat{=}\,}
\newcommand{\eqon}[1]{\stackrel{{#1}}=}
\newcommand{\eqonN}{\,\eqon{\sscr\cN}\,}
\newcommand{\pb}[1]{\stackrel{{#1}}=}
\usepackage{pgf}
\makeatletter
\newcommand*{\pgfunderleftarrow}{%
  \@ifstar
    {\let\ifpgf@depth\iftrue\mathpalette\@pgfunderleftarrow}
    {\let\ifpgf@depth\iffalse\mathpalette\@pgfunderleftarrow}%
}
\newcommand*{\@pgfunderleftarrow}[2]{%
  #2%
  \edef\pgf@math@fam{\the\fam}%
  \pgfpicture
    \pgfsetbaseline{0pt}
    \pgf@relevantforpicturesizefalse      
    \pgfsetroundcap                       
    \pgfsetarrowsend{to}
    \pgfutil@tempdima=0.28pt%
    \advance\pgfutil@tempdima by.8\pgflinewidth%
    \pgfutil@tempdima-4\pgfutil@tempdima
    \sbox\pgfutil@tempboxa{$\m@th\fam\pgf@math@fam#1#2$}%
    \advance\pgfutil@tempdima-\dp\pgfutil@tempboxa
    \pgfutil@tempdimb\wd\pgfutil@tempboxa
    \pgfpathmoveto{\pgfqpoint{0pt}{\pgfutil@tempdima}}%
    \pgfpathlineto{\pgfqpoint{-\pgfutil@tempdimb}{\pgfutil@tempdima}}%
    \pgfusepath{stroke}
    \ifpgf@depth
      \pgf@relevantforpicturesizetrue
      \pgfpathmoveto{\pgfqpoint{0pt}{-\pgfutil@tempdimb}}%
      \pgfusepath{use as bounding box}%
    \fi
  \endpgfpicture
}
\makeatother
\newcommand{\pbi}[1]{\pgfunderleftarrow{#1}}

\newcommand{\sscr}{\scriptscriptstyle\rm}

\usepackage{manfnt}
\reversemarginpar

\newcommand{\ella}{l_o}
\newcommand{\xiext}{\bar\xi^\Phi}

\newcommand{\Diff}{{\rm{Diff}}}

\newcommand{\ba}{{\hat\b}}
\newcommand{\pbg}{\smash{\underset{\raisebox{.7ex}[0pt][0pt]{\footnotesize{$\leftarrow$}}}{\g}{}_{\m\n}}}

\DeclareFontFamily{U}{matha}{\hyphenchar\font45}
\DeclareFontShape{U}{matha}{m}{n}{
      <5> <6> <7> <8> <9> <10> gen * matha
      <10.95> matha10 <12> <14.4> <17.28> <20.74> <24.88> matha12
      }{}
\DeclareSymbolFont{matha}{U}{matha}{m}{n}
\DeclareMathSymbol{\oright}       {2}{matha}{"69}

\usepackage{pifont}
\newcommand{\cmark}{\ding{51}}%
\newcommand{\xmark}{\ding{55}}%

\begin{document}

\title{\bf General gravitational charges on null hypersurfaces}

\author{\Large{Gloria Odak, Antoine Rignon-Bret and Simone Speziale}
\smallskip \\ 
\small{\it{Aix Marseille Univ., Univ. de Toulon, CNRS, CPT, UMR 7332, 13288 Marseille, France}} }
\date{November 27, 2023}

\maketitle

\begin{abstract}
We perform a detailed study of the covariance properties of the symplectic potential of general relativity on a null hypersurface, and of the different polarizations that can be used to study conservative as well as leaky boundary conditions. This allows us to identify a one-parameter family of covariant symplectic potentials. 
We compute the charges and fluxes for the most general phase space with arbitrary variations. 
We study five symmetry groups that arise when different restrictions on the variations are included. 
Requiring stationarity as in the original Wald-Zoupas prescription selects a unique member of the family of symplectic potentials, the one of Chandrasekaran, Flanagan and Prabhu.
The associated charges are all conserved on non-expanding horizons, but not on flat spacetime. We show that it is possible to require a weaker notion of stationarity which selects another symplectic potential, again in a unique way, and whose charges are conserved on both non-expanding horizons and flat light-cones.
Furthermore, the flux of future-pointing diffeomorphisms at leading-order around an outgoing flat light-cone is positive and reproduces a tidal heating 
plus a memory term. 
We also study the conformal conservative boundary conditions suggested by the alternative polarization and identify under which conditions they define a non-ambiguous variational principle.
Our results have applications for dynamical notions of entropy, and are useful to clarify the interplay between different boundary conditions, charge prescriptions, and symmetry groups that can be associated with a null boundary. 

\end{abstract}

\tableofcontents

\section{Introduction}

Chandrasekaran, Flanagan and Prabhu (CFP) characterized the symmetry group of general relativity on generic null hypersurfaces as an extension of the BMS group to include arbitrary diffeomorphisms and Weyl transformations of any 2d space-like cross-section \cite{Chandrasekaran:2018aop}.\footnote{The same group that can be obtained at null infinity relaxing the fall-off conditions in a way compatible with renormalization of the symplectic potential \cite{Freidel:2021yqe}, and was in that context referred to as BMSW group. See also \cite{Flanagan:2023jio}.} They then used the Wald-Zoupas (WZ) procedure \cite{Wald:1999wa} to prescribe charges for this symmetry group and study their flux-balance laws. The charges they obtained satisfy important properties, and are conserved on shear-free and expansion-free hypersurfaces, or equivalently non-expanding horizons (NEHs) in vacuum.
The analysis was then specialized to NEHs and further advanced in \cite{Ashtekar:2021kqj}, explaining for instance how the NEH area is the charge aspect associated with the global sector of the Weyl transformations. The same Weyl transformations play also a prominent role in the investigation of possible black hole soft hairs by Hawking, Perry and Strominger \cite{Hawking:2016sgy}.

The CFP construction is based on the covariant phase space with a specific choice of symplectic potential, associated with Dirichlet boundary conditions, and on a specific set of restrictions of the variations, corresponding to a certain universal structure constructed along the guidelines of what has successfully been done at future null infinity (see e.g. \cite{Ashtekar:2014zsa} for a review). These restrictions clarify in particular the issue of ambiguous null boundary terms that was raised in \cite{Lehner:2016vdi}. It was then pointed out in \cite{Chandrasekaran:2021hxc} that the CFP charges differ from what one would obtain following a procedure \`a la Brown-York by a term generated by an anomalous transformation under diffeomorphisms, thus bringing to the forefront the extension of the covariant phase space to include anomalies described in \cite{Chandrasekaran:2020wwn,Freidel:2021cjp} (see also \cite{Chandrasekaran:2021vyu,Odak:2022ndm}).
We show here that the anomaly term in the charges is crucial to make them covariant, and explain why this seemingly counterintuitive statement is actually natural.

A shortcoming of the CFP charges is that some of them are not conserved even in the absence of radiation, specifically those whose aspect is the area, since this grows on a flat light-cone.
This limits their applicability to study physical processes, for instance a spherical collapse would see a flat light-cone bend into an event horizon, and this process will be poorly described by a charge that is already varying prior to any matter infalling.
This raises the question whether a different prescription for the charges exists that is free of this shortcoming. Indeed, the Noether charges as well as the Wald-Zoupas charges are not guaranteed to be unique, and may depend on a choice of polarization made in writing down the symplectic potential \cite{Iyer:1995kg,Harlow:2019yfa,Freidel:2020xyx,Freidel:2021cjp,Odak:2021axr,Chandrasekaran:2021vyu,Odak:2022ndm}. 
We show that there exists a different choice of polarization that leads to charges which are conserved on flat light-cones, as well as on shear-free and expansion-free hypersurfaces. 
The polarization we use was previously considered in \cite{Hopfmuller:2018fni}, although with a restriction that spoiled its covariance.
The application of the new charges 
to study dynamical processes of BH formation was anticipated in \cite{Rignon-Bret:2023fjq}, see also \cite{WaldZhang} on this.
It is the unique one that satisfies the Wald-Zoupas covariance condition and a stationarity condition interpreted in a weaker sense than the original WZ paper and that allows one to include both shear and expansion-free surfaces and flat light-cones. A further useful property of this potential is that
the flux of future-pointing diffeomorphisms at leading-order around an outgoing flat light-cone is positive, and reproduces the tidal heating term plus a memory term.

We also show that this polarization leads to conformal boundary conditions on a null hypersurface that provide an alternative resolution to the boundary term ambiguities of \cite{Lehner:2016vdi} based on Dirichlet boundary conditions, and discuss the residual ambiguities that would be present in the corner terms.

One of the restrictions on the variations considered in the CFP paper concerns the inaffinity of the normal to the null hypersurface. This restriction plays an important role. Relaxing it reintroduces the ambiguities of \cite{Lehner:2016vdi} and prevents a complete implementation of the WZ procedure. Nonetheless, the authors considered the possibility that this variation may be physically relevant, and the question was left open whether one could construct WZ charges in this larger phase space. 
We investigate this issue with a general analysis of the covariant phase space in which all restrictions on the variations are removed one by one, until we are left with the minimal condition that the boundary is preserved. Removing restrictions the symmetry group gets enlarged from the CFP group to a group which includes super-translations of arbitrary time-dependence, to the complete hypersurface diffeomorphisms, to an extension of the hypersurface diffeomorphism with one additional free function which applies to the largest phase space.

Having identified the symmetry groups, we look at the symplectic potential for all phase spaces. 
We perform a systematic calculation of all anomalies and provide their interpretation. We identify a one-parameter family of symplectic potentials that satisfy the covariance condition in the larger phase spaces, family that includes the conformal polarization. However, none of them satisfies the stationarity condition, neither in the original sense nor in the weaker sense. As a consequence, the new symmetries produce fluxes which don't vanish in stationary solutions such as non-expanding horizons and flat light-cones. So while it is interesting to notice that this extension of the phase space is possible, it remains unclear to us how it should be used in physical applications.

The stationarity condition can instead be satisfied for the whole family if the variation of the inaffinity is non-zero but fixed to be proportional to the variation of the expansion. The proportionality parameter labels the members of the family, and when it vanishes the CFP choice is recovered.
In all other cases one finds symmetry vector fields with a non-trivial metric dependence, and will be investigated elsewhere. 

We complete the analysis providing the general expression for the Noether charges for arbitrary variations and a two-parameter family of polarization that include the covariant ones. The covariant ones are the only ones that are invariant under arbitrary choices such as rescaling of the null normal, change of embeddings, and change of rigging vector. We explain under which conditions these charges satisfy the Wald-Zoupas prescription.

Our results strengthen on the one hand  the value of the universal structure defined in \cite{Chandrasekaran:2018aop}, and  enrich it proposing an alternative symplectic potential with improved stationary properties. On the other hand they open the possibility of working with a much weaker universal structure while preserving the Wald-Zoupas covariance criterium, if the lack of stationarity can be dealt with.

At the technical level, a difference between our approach and 
\cite{Chandrasekaran:2018aop} is that we take a spacetime description as starting point, as opposed to the intrinsic description based on hypersurface tensors. Our complementary approach can be useful to provide a new angle on their analysis, and allows one to relate it to the Newman-Penrose (NP) formalism. This gives us useful tools for the systematic analysis of anomalies in the phase space. For instance, the geometry of a null hypersurface depends only on the equivalence class of normals under rescaling. But many quantities that appear in the phase space depend also on a specific choice of normal representative, as well as on an auxiliary rigging vector used to define a local projector on space-like cross-sections. Identifying the quantities which are independent of these two auxiliary and non-geometric characteristics is straightforward using internal transformations of the NP tetrad which have long since been tabulated, and are for instance denoted class-I and class-III in \cite{Chandra}. One of our technical results is to point out the direct link between lack of invariance under these transformations and anomalies in the covariant phase space. 
Another technical difference is that we don't use the notation for the covariant phase space \`a la Wald, where field variations are interpreted as tangent vectors, but as in  as differential forms \cite{Donnelly:2016auv,Harlow:2019yfa,Freidel:2021cjp}. 
This makes it easier to distinguish the Lie derivative in field space from those in spacetime, which is the basic step to study anomalies. Some of the details we present are therefore mere translations of the results of \cite{Chandrasekaran:2018aop} in the formalism based on spacetime tensors and on the exterior calculus notation for the covariant phase space. This makes for a paper longer than initially intended, but we hope that the dictionary it provides will be of use to navigate the literature.

\bigskip

As this paper was being completed we learned of a similar analysis by Venkatesa Chandrasekaran and Eanna Flanagan which has considerable overlap with our results and which is being submitted simultaneously to the arXiv \cite{Chandrasekaran:2023vzb}.

\bigskip

We use mostly-plus spacetime signature. Greek letters are spacetime indices, and we will sometimes denote scalar products by a dot. 
When needed, lower case latin letters $a,b,...$ are hypersurface indices, and upper case latin letters $A,B,...$ are indices for  2d cross-sections of the hypersurface.
In all cases, $(,)$ denotes symmetrization, $\la,\ra$ trace-free symmetrization, and $[,]$ antisymmetrization. An arrow under a $p$-form means pull-back on $\cN$, $\eqons$ means on-shell of the field equations, and $\eqonN$ means an equality valid at the null boundary only. 
We use units $16\pi G=c=1$. 

\section{Null hypersurfaces}\label{SecNull}

In this Section we review basic facts of null hypersurfaces. This is useful to fix the notation, but it will also allows us to highlight properties that are often scattered between different literature, and to provide somewhat of a dictionary. 
 We will in particular review the distinction between intrinsic and extrinsic geometries, using spacetime covariant notation and offering the translation to hypersurface indices on the one hand and to Newman-Penrose (NP) notation on the other hand. We will recall the notion of class-III and class-I invariance to talk about quantities which are  independent respectively of the choice of normal and of rigging vector.
We will then recall Sachs' identification of constraint-free data and how it allows one a clear distinction between conservative and radiative or leaky boundary conditions,
and finally some useful expressions that arise when working with the special coordinate system provided by affine coordinates.

We consider a null hypersurface $\cN$ defined by a cartesian equation $\Phi=0$, and denote its normal
\be\label{defl}
l_\m:\eqonN -f\p_\m \Phi, \qquad l^2{\eqonN}0,
\ee
with $f>0$ as to have the vector future-pointing. The corresponding vector $l^\m$ is null and hypersurface orthogonal, hence it is also 
tangent to $\cN$ and geodetic,
\be\label{lgeo}
l^\m\na_\m l^\n{\eqonN}k l^\n.
\ee
The hypersurface is thus naturally fibrated by null geodesics, and $k=0$ if they are affinely parametrized.
The chosen tangent vector is referred to as generator of $\cN$. 
We will assume that $\cN$ has topology $I\times S$, where $S=S^2$ and $I$ is some interval in $\R$. 
If $I=\R$ in affine coordinates the hypersurface is called complete, and in this case all null geodesics extend indefinitely in both directions. It is called semi-complete if it extends indefinitely in one direction only, and has a boundary in the other direction caused for instance by the formation of caustics or  crossings. 
We will often use adapted coordinates $(\Phi,x^a)$, where $x^a$, $a=1,2,3$ are coordinates on the leaves of the $\Phi$ foliation. The condition that $\cN$ is null then induces a partial gauge-fixing of the metric given by
\be
g^{\Phi\Phi}:=g^{\m\n}\p_\m\Phi\p_\n\Phi{\eqonN}0.
\ee

For space-like and time-like hypersurfaces, there is a canonical choice of normal with unit norm. This makes the normal independent of the embedding of the hypersurface,
namely invariant under a change of parametrization $\Phi\mapsto \Phi'=\Phi F(x)$ with $F$ smooth at $\cN$  that preserve the location of the boundary. 
No such preferred choice exists in the null case.
The function $f$ is thus arbitrary, and one has to check on a case by case basis whether a given quantity is independent of the embedding or not.
The geometry of the hypersurface is on the other hand only sensitive to the equivalence class $[l=Al]$ of normals identified up to an arbitrary rescaling.
This rescaling can be obtained in two independent ways. First, changing the choice of $f$. Second, changing the embedding, which has the effect of multiplying $f$ by $F$.

For instance, the inaffinity $k$ appearing in \eqref{lgeo} is not a geometric property of a generic null hypersurface, since it depends on $f$ and $\Phi$.
An explicit calculation using \eqref{defl} gives in fact
 \be\label{kmoche}
 k=  \pounds_l \ln f -\f f2 \p_\Phi g^{\Phi\Phi},
 \ee
written in adapted coordinates. 
If follows that $k$ depends on the transversal derivative of the metric, namely it contains information about the extrinsic geometry.
We can rewrite \eqref{kmoche} in a more covariant form if we parametrize an arbitrary extension of the normal as $l=-fd\Phi +\Phi\, v$, then
 \be\label{kgen}
 k=  \pounds_l \ln f - \f12 \pounds_n l^2 + \f1f l\cdot v,
 \ee
where $n$ is any null vector such that $n\cdot l=-1$. This expression is slightly misleading because it may give the impression that $k$ depends on $v$, but this dependence cancels out between the second and third term to give back \eqref{kmoche}.\footnote{Independence of $k$ from the extension of $l$ can be also checked showing that $l'=l+\Phi v$ gives the same $k$ as $l$, and it means that different $k$ imply different $f$ at fixed spacetime metric.}
From this general expression one can also read off the special values when 
the extension is null everywhere ($l^2=0$, $v\neq 0$), when it is  hypersurface-orthogonal everywhere ($v= 0$), or both.
A typical example of the first special case is Kerr's principal null direction, hypersurface orthogonal only at the horizon. For the second special case, $l$ is normal to a foliation that has a single null leaf, and for the third, $l$ is normal to a null foliation.

This discussion is valid for a generic null hypersurface. 
If it is a Killing horizon in a spacetime with a global isometry, then the possibility of selecting as generator the preferred Killing vector of asymptotic unit-norm allows one to eliminate these arbitrariness and interpret $k$ as surface gravity. 
If it is a non-expanding horizon, 
 the arbitrariness in the normal can be reduced to constant rescaling. Every member of the equivalence class has $k=0$ (not to be confused with the surface gravity of the NEH, which is the arbitrary parameter of the Weyl rescaling symmetry vector field). We recall that the definition of a NEH coincides with a shear-free and expansion-free null hypersurface in vacuum, but it is slightly more specific in the presence of matter, where it requires the stronger condition $R_{\m\n}l^\m \eqonN \a l_\n$ for some function $\a$, as opposed to $R_{\m\n}l^\m l^\n \eqonN 0$ satisfied by a shear-free and expansion-free null hypersurface. In this paper we will only deal with vacuum general relativity, hence we will use NEH as a synonym of a shear and expansion-free hypersurface.

The null vector $n$ introduced in \eqref{kgen} is known as the rigging vector, and it is a convenient tool to work on null hypersurfaces. 
It allows one the use of covariant expressions at all times and to avoid hypersurface indices, thus making the relation to spacetime objects transparent. It also allows one to use the Newman-Penrose (NP) formalism and the numerous results that have been derived in that language. To that end, we complete the pair $(l,n)$ to a doubly-null NP tetrad $(l,n,m,\bar m)$ on $\cN$. 

The downside of the rigging vector approach is its reliance on an arbitrary choice of auxiliary vector. But it is quite easy to check which quantities are independent of this choice. The arbitrariness is a 2-parameter family given by 
\be\label{classI}
n\rightarrow n+\bar a m+a\bar m+|a|^2l, \qquad m\rightarrow m+al, \qquad a\in\C.
\ee
Quantities which are invariant under \eqref{classI} are independent of the choice of auxiliary rigging vector. For instance, it is easy to check that \eqref{kgen} is invariant.
The map \eqref{classI} is an internal Lorentz transformations of the NP tetrad that corresponds to the two translations of the ISO(2) little group stabilizing $l$. 
We will refer to it as a class-I transformation (of the NP tetrad), following \cite{Chandra}.
In this classification, class-II transformations are the two null translations of the ISO(2) little group stabilizing $n$. They change $l$ and disalign it from the normal to the hypersurface, and will not be considered in the rest of the paper.
The remaining two internal transformations are the class-III spin-boost transformations 
\be\label{classIII}
(l,n,m,\bar m)\rightarrow (Al,A^{-1}n,e^{i\varphi}m,e^{-i\varphi}\bar m).
\ee 
The boost transformation acts as a rescaling of the normal by an arbitrary real function $A$. 
Therefore, quantities invariant under this boost are independent of the choices of $f$ and of the embedding used when writing \eqref{defl}. 
This is not the case for the inaffinity, which transforms as $k\to A(k+\pounds_l A)$.

An important result of \cite{Chandrasekaran:2018aop} is that the equivalence class
\be\label{CFPclass}
[l, k]=[Al, A(k+\pounds_l A)]
\ee
can be taken to be the universal background structure in the covariant phase space of metrics with a null hypersurface.\footnote{This is referred to as universal boundary structure. We prefer the adjective background to emphasize that these are quantities that will not be varied in the phase space. The paper \cite{Chandrasekaran:2018aop} also introduces a notion of `universal intrinsic structure', based on purely intrinsic quantities, and which we do not consider here. } 
That is, any two metrics with a null hypersurface admit a coordinate system in which they have the same \eqref{CFPclass}.
Elements in the universal structure \eqref{CFPclass} must thus be class-III invariant.

These internal Lorentz transformations are thus practical tools to discern the quantities that depend solely on the geometry of $\cN$ from those that depend on additional structures or choices. To reiterate, class-I invariance means independence from the choice of rigging vector,\footnote{\label{FootCarroll} The extended structure of a null hypersurface plus a specific choice of normal is called Carollian structure in some literature, and the further extension including a specific choice of rigging vector a `ruled' or rigged Carollian structure, see e.g. \cite{Ciambelli:2019lap}.} and class-III invariance guarantees independence from rescaling the normal, namely from the choice of $f$ and from reparametrizations $\Phi\mapsto \Phi'=\Phi F(x)$.  One theme of this paper will be that lack of class I and class III invariance translates to anomalies in the covariant phase space.

If need be to select a specific rigging vector, there are two natural ways to do so that are common in the literature. The first is to require it to be parallel-transported along $l$ on $\cN$, see e.g. \cite{Newman:1962cia}. This choice is unique, and fixes the class-I transformation so that the NP spin-coefficient $\pi$ vanishes.\footnote{Requiring $(m,\bar m)$ to be also parallel-transported will further fix the spin part of the class-III so that the NP coefficient $\eps$ is real. This is the same letter used below for the volume form, but being the first a scalar and the second a form no confusion should hopefully arise.} 
The second way is to require it to be adapted to a given $2+1$ foliation of $\cN$. This choice is again unique once the foliation is given.
In this case the class-I transformation is fixed setting to zero two of the three components of the pull-back of $n$, thus making it hypersurface orthogonal within $\cN$.
It follows that the 2d planes spanned by $(m,\bar m)$ integrate to the leaves of the foliation.

A volume form on $\cN$ can be defined from the spacetime volume form $\eps$ via
\be
\eps= - l\w\eps_\cN.
\ee
The conventional minus sign here follows from assuming $l$ outgoing, and would be plus if incoming. 
The volume form $\eps_\cN$ is class-I invariant but not class-III invariant because it depends on $f$.
This formula defines actually an equivalence class of volume forms, related by adding any 3-form containing $l$. A convenient representative of this equivalence class can be chosen using the rigging vector as
\be\label{epsN2}
\eps_\cN:=\pbi{i_n \eps} = \f{\sqrt{-g}}{f}d^3x,
\ee
where the second equality uses adapted coordinates $(\Phi,x^a)$, and the arrow under the form means pull-back on $\cN$.
We will make this choice from now on. Written in this way, class-I invariance may not appear as obvious, but it follows from the pull-back and the fact that $m^\m$ is tangent to $\cN$.

On $\cN$, we also define the space-like area form
\be\label{defepsS}
\eps_S:= i_l\eps_{\cN}=i\pbi{m\w \bar m}, \qquad i_l \eps_S=0, \qquad \qquad \eps_\cN= -n\w \eps_S.
\ee
It is class-I invariant and defined independently of any choice of foliation of $\cN$.
It satisfies
\be\label{des}
d\eps_S = \th\eps_\cN,
\ee
where $\th$ is the expansion of $l$, as defined below. Notice that $\eps_S$ so defined can contain components along the null direction, 
even if $n$ is adapted to a foliation and $(m,\bar m)$ are integrable. Choosing affine coordinates eliminates these components.
From this equation one also derives the following useful identity,
\be\label{anvedi}
(\pounds_l +\th)X \, \eps_\cN = d(X\eps_S).
\ee

The rigging vector is also handy to introduce a local projector on 2d space-like planes, given by
\be
\g_{\m\n}:=g_{\m\n}+2l_{(\m}n_{\n)}=2m_{(\m}\bar m_{\n)}.
\ee
Its pull-back $\pbg$, or $\g_{ab}$ in hypersurface indices, coincides with the pull-back of the spacetime metric.
This is the (degenerate) induced metric, whose null direction is given by $l^\m$ itself. 
The class-III invariant pair 
$(\pbg,l^\m\eps_\cN)$ contains six independent quantities, which
are the analogue of the induced geometry in the non-degenerate case.

For the extrinsic geometry, we look at the pull-back of the gradient of the normal vector. This quantity gives the extrinsic curvature in the case of a space-like or time-like hypersurface. In the null case, $W_{\m}{}^\n:=\na_{\underset{\leftarrow}\m} l^\n $ defines a a purely hypersurface objet, 
satisfying $n^\m W_\m{}^\n=0=W_\m{}^\n l_\n$, and $l^\m W_\m{}^\n=k l^\n$. It is related to the Weingarten map, which is the reason for the notation $W$. The actual map is given using hypersurface indices as in \cite{Chandrasekaran:2018aop,Chandrasekaran:2021hxc}, but that definition is equivalent to ours in terms of covariant 4d indices. To see the geometric content of this map, it is convenient to use the rigging vector and decompose it as follows,
\begin{align}\label{defW}
W_{\m}{}^\n:=\na_{\underset{\leftarrow}\m} l^\n & \stackrel{\cN}{=} \om_{\pbi{\m}}l^\n + \g^{\n}_{\r}B_{\pbi{\m}}{}^{\r}
\\\nn &= \big((\bar \a +\b)\bar m_{\pbi{\m}}-\eps n_{\pbi{\m}} \big) l^\n - (\s \bar m_{\pbi{\m}} + \r m_{\pbi{\m}}) \bar m^\n +{\rm cc}.
\end{align}
The second line makes reference to the NP formalism (with mostly-plus signature, we use the conventions of \cite{Ashtekar:2000hw}\footnote{This formula can be found for example in references on the NP formalism (e.g. \cite{Chandra}). The NP formalism assumes an extension of $l$ which is null everywhere, but since the derivative index is here pulled-back on $\cN$, it is valid for an arbitrary extension as well.}), and the various tensors there appearing are:
\begin{subequations}\label{Wdec}\begin{align}\label{defB}
& B_{\m\n}:=\g_\m^\r\g_\n^\s \na_\r l_\s = \f12 \g_\m^\r\g_\n^\s \pounds_l \g_{\r\s} {\eqonN}  \s_{\m\n}+\f12\g_{\m\n}\th, \\
& \s_{\m\n}:=\g_{\la \m}^\r\g_{\n\ra}^\s \na_\r l_\s = -\bar m_\m\bar m_\n \s+cc,
\qquad\ \ \th:=2m^{(\m}\bar m^{\n)}\na_\m l_\n =-2\r, \\
& \om_{{\m}}:=- \eta_{\m}-kn_\m, \qquad\quad \eta_{\m}:=\g_\m^\r n^\s\na_\r l_\s=-(\a+\bar\b)m_\m+cc, 
\qquad l^\m\om_\m=k=2\re(\eps). \label{defdot}
\end{align}\end{subequations}
Here $B$ is the deformation tensor, whose antisymmetric part vanishes because $l$ is hypersurface orthogonal at $\cN$, $\s$ is the shear and 
$\th$ the expansion; $\om$ is the rotational 1-form of isolated and non-expanding horizons \cite{Ashtekar:2000hw,Ashtekar:2004cn}, satisfying $\om\cdot l=k$; $\eta$ is 
the connection 1-form on the normal time-like planes spanned by $(l,n)$, 
whereas the complementary quantity $\a-\bar\b$ is the 2-sphere connection of the covariant derivative $\eth$ used in NP calculus \cite{NP62,Booth:2006bn}. $\eta$ is sometimes called Hajicek 1-form \cite{hajivcek1973exact}, or twist, since it is related to the non-integrability of the normal planes via
\be\label{nltoeta}
\g_{\m\n} [n,l]^\n= \eta_\m - \g_{\m}^{\n}( l^\r\na_\r n_\n - \p_\n\ln f).
\ee

The Weingarten map depends on a specific choice of normal and not on the equivalence class. 
It is nonetheless useful to describe the geometry of the null hypersurface.
From \eqref{defB}, we see that the shear and expansion are entirely determined by the induced metric and a choice of $l$, so they are part of the intrinsic geometry. 
The dependence on the scaling of $l$ can be eliminated if we look at the densitized expressions $\s\eps_\cN$ and $\th\eps_\cN$ which are class-III invariant.

Transversal derivatives of the metric enter the inaffinity $k$ and the twist $\eta_\m$. These quantities could be taken as the analogue of the extrinsic geometry, but they are ambiguous since they depend on the choice of $l$ representative and not on the equivalence class. 
This dependence can be partially removed if we consider the following shifts, 
\begin{align}
&	\bar{k} := k - l^\m \p_\m \ln{f} = -\f f2 \p_\Phi g^{\Phi\Phi}, \label{class3invariantk} \\
	\label{nltoeta1}
& \bar\eta_\m:=\eta_\m +\g^\n_\m\p_\n\ln f = \g_{\m\n}([n,l]^\n + l^\r\na_\r n^\n) = m_\m (\bar m_\n [n,l]^\n + \pi) +{\rm cc},
\end{align}
where $\pi$ here is one of the NP coefficients.
$\bar \eta_\m$ and $\bar k\eps_\cN$ are invariant under changes of $f$, but not under changes of embedding $\Phi\to\Phi F(x)$. Therefore they are still not class-III invariant, but at least satisfy the weaker requirement of being 
independent of the choice of normal representative at fixed embedding.\footnote{This is consistent with the statement in \cite{Hopfmuller:2016scf} that a quantity like $\bar k\eps_\cN$ here is class-III boost invariant, because that paper works with a fixed 2+2 foliation.}
If we keep the embedding fixed, $\bar k\eps_\cN$ is fully unambiguous. 
However $\bar\eta_\m$ is not, because it inherits from $\eta_\m$ a dependence on the rigging vector, hence it is still not a genuine measure of the extrinsic geometry of $\cN$.
In fact, even though  the Weingarten map is independent of the choice of rigging vector, the decomposition we used on the right-hand side of \eqref{defW} introduces a dependence on it: only $\th$, $k$ and (the scalar contraction) $\s$ are class-I invariant, whereas $\s_{\m\n}$, $\eta_\m$ and $\om_\m$ are not. 
For convenience, the transformation properties of all quantities are summarized in Table~\ref{invTable}, with the details reported in Appendix~\ref{AppLorentz}.

The only case in which (the pull-back of) $\bar \eta_\m$ is class-I invariant is on a non-expanding horizon with $k=0$. And in fact it characterizes the shape of a non-expanding horizon via the Noether charge construction \cite{Ashtekar:2021kqj}. To use it as a measure of the extrinsic geometry of a general $\cN$, one has to fix the class-I gauge freedom. If we do so taking $n$ parallel transported by $l$ the NP spin coefficient $\pi$ vanishes and can identify the twist $\bar \eta_\m$ (or equivalently $\eta_\m$ with a gradient normal representative) with the non-integrability of the time-like planes, thanks to \eqref{nltoeta1}. Below we will however find it more convenient to fix instead $n$ to be adapted to a foliation of $\cN$, and we will then show that $\bar\eta_\m$
determines the evolving Noether charges associated with the leaves of that foliation.

\begin{table}
\begin{center}\begin{tabular}{c|cc|cc} quantity & Rigging-vector & Rescaling  & Boost & Spin   \\ 
 &  independence &  independence &  weight &  weight \\ \hline
$\s_{\m\n}$ & \xmark & \xmark & 1 & 0  \\
$\s$ & \cmark & \xmark & 1 & 2  \\
$\th$ & \cmark&\xmark & 1 & 0  \\
$\eps_\cN$ &\cmark&  \xmark & $-1$ & 0  \\
$k$ & \cmark &  \xmark & 1+inhom & 0  \\
$\eta_\m$ & \xmark & \xmark & 0+inhom & 0 \\
$\a+\bar\b$ & \xmark & \xmark & 0+inhom & -1 \\
$\g_{\m\n}[l,n]^\n$ & \xmark & \cmark &  0 & 0
\end{tabular}\end{center}
\caption{\label{invTable} \small{\emph{Behaviour under class-I and class-III transformations. Quantities that are not invariant under \eqref{classIII} can be characterized in terms of their boost and spin weights, respectively
$a$ and $b$, defined by 
$X\to A^a e^{ib\varphi}X$ (up to possible inhomogeneous terms) under \eqref{classIII}. 
The boost weight can also be interpreted as a conformal weight, for instance in the case of future null infinity where the normal is the gradient of the conformal rescaling of the metric.
}}}
\end{table}
We conclude with two more remarks about the Weingarten map. First, its trace is given by
\be\label{Wtrace}
W:=W_{\pbi\m}{}^\m =\na_\m l^\m +\f12\p_n l^2= \th+k, 
\ee
and provides the boundary term for the variational principle with Dirichlet boundary conditions on a null hypersurface \cite{Parattu:2015gga,Lehner:2016vdi,Jubb:2016qzt,Oliveri:2019gvm}, the equivalent of the Gibbons-Hawking-York term.
The discrepancy between the trace of the Weingarten map and the divergence of the normal may look unfamiliar, but it would occur also  in the time-like case if the normal $\t$ is not of unit-norm off the hypersurface: 
$K= \na_\m \t^\m +\f12\p_\t \t^2$, where $K_{\m\n}:=q_\m^\r \na_\r \t_\n$.

Second, an alternative covariant construction of the Weingarten map can be given in terms of the ``half-projector" $\Pi_\m{}^\n := \g_\m^\n -n_\m l^\n$, defining
$
\tl W_\m{}^\n:=\Pi_\m{}^\r \na_\r l^\n$. This tensor is rigging-vector dependent, but not its pull-back on the hypersurface.
This pull-back is the definition used in \cite{Chandrasekaran:2021hxc}, and coincides with \eqref{defW}. The trace also coincides with \eqref{Wtrace}, namely
$\tl W_\m{}^\m=W.$

\subsection{Foliations} 

The volume form $\eps_\cN$ is not class-III invariant, and depends on the full spacetime metric determinant $\sqrt{-g}$. On non-degenerate hypersurfaces choosing a unit-norm normal makes the volume form depend only on the determinant of the  induced metric.
The unit-norm option does not exist in the null case, but one can achieve a similar result introducing a $2+1$ foliation of $\cN$.
The foliation can be arbitrary, provided that its leaves are space-like. 
We take it to be defined by the level sets of some scalar function $\l$, and denote $x^a=(\l,x^A)$ the coordinates adapted to it.

Note that if we take this choice together with the foliation defined by $\Phi$, we obtain spacetime a coordinate system $(\Phi,\l,x^A)$ adapted to a 2+2 foliation of spacetime (see e.g. \cite{dInverno:1980kaa}). Our choice of letters for these coordinates 
is meant to preserve generality of the formalism with respect to common applications. For example, 
to make the link the Schwarzschild metric in retarded Bondi coordinates we would take $(\l,\Phi)=(u,r-2M)$ and $\cN$ is the white hole horizon, or using advanced time instead  $(\l,\Phi)=(v,2M-r)$ and $\cN$ is the black hole horizon. 
We can also keep assuming $l^\m$ future pointing namely $g^{\Phi\l}<0$ without loss of generality. In the first case this leads to $g^{ur}<0$, in the second case to 
$g^{vr}>0$.
Or if $\cN$ is a null cone in Minkowski in a doubly-null foliation, we can identify $\Phi=u:=t-r$ and $\l=v:=t+r$.
Since $\l$ is a (null) time, we will refer to $\p_\l$ as a time derivative, and  use a dot to indicate it.

 In these coordinates, 
\be
\sqrt{-g} = -\f1{g^{\Phi \l}}\sqrt\g,
\ee
where $\g$ is the determinant of the space-like metric $\g_{AB}$ on the 2d leaves. Hence, 
\be
\eps_\cN = \f{\sqrt{\g}}{l^\l} d\l d^2x.
\ee 
We see that it is a completely intrinsic quantity, but it is still not class-III invariant and contains more information than the 2d area form $\g$: it depends also on the extent of $l$ via $l^\l$.
If we now choose $f=-1/g^{\Phi\l}$, 
we obtain $l^\l=1$ and
\be\label{epsgamma}
\eps_\cN=\sqrt{\g}d\l d^2x.
\ee
Notice that $\l$ does not need to be a parameter along the null geodesics. In general after making these choices, 
\be\label{lgen}
l^a= (1,-b^A), \qquad \pbi{g_{ab}}=\mat{\g_{AB}b^Ab^B}{\g_{AB}b^B}{}{\g_{AB}}, 
\ee
and the vector $b^A$ acts as a shift vector for the 2+1 foliation defined by $\l$. 
If we partially fix the coordinate gauge requiring that $x^A$ are conserved along the generators, then we are setting the shift vector to zero, and $l^a=(1,0,0)$.
In terms of the spacetime metric, this partial gauge-fixing reads $g^{\Phi A}\eqonN0$. We refer to it as partial Bondi gauge, as in \cite{DePaoli:2017sar,Geiller:2022vto}.
The foliation-dependent choice $f=-1/g^{\Phi \l}$ 
was referred to as `canonical' normalization in \cite{Oliveri:2019gvm}, for its analogy with the ADM space-like case, since 
$1/g^{\Phi \l}$ plays the role of lapse in the $3+1$ decomposition with null slices \cite{Alexandrov:2014rta}. 

The simplification \eqref{epsgamma} gives to the volume form a similar structure to the one of non-degenerate hypersurfaces 
(albeit in terms of a codimension-2 determinant), and it is often used in the literature, e.g. \cite{Poisson:2009pwt}. It is valid only in the foliation chosen,
but in the partial Bondi gauge it remains valid for any new foliation obtained by a super-translation $\l'=\l+T(x^A)$. 
We will however not do this choice in the following, neither for $f$ nor for the coordinates, and keep fully general $f$ and $\eps_\cN$.

\bigskip

A common choice of $2+1$ foliation is the one induced by the intersections of $\cN$ with a space-like foliation. In this case the cross-sections of $\cN$ provide the boundary $\p\Si$ of each 3d space-like leaf $\Si$. Let us denote by $\t$ the unit-norm normal to the space-like foliation, and parametrize the scalar product as follows, \be\label{ldottau}
l\cdot\t {\eqonN} -\f1{\sqrt{2}}e^{-\ba}.
\ee
The overall minus sign is due to the fact that both vectors are future pointing. The quantity $\ba$ has no geometric meaning per se, since it is not class-III invariant. It can be used to measure the change of geometric tilt between $\cN$ and $\Si$ only if $l$ is kept fixed.
The unit-norm normal to the cross-section within $T\Si$ is
\be
\hat r^\m{\eqonN} \pm \sqrt 2 e^\ba q^\m_\n l^\n, \qquad q_{\m\n}:=g_{\m\n}+\t_\m\t_\n,
\ee
where the sign is plus if $\cN$ is the outgoing null hypersurface from the boundary of $\Si$, and minus if it is the incoming one.
It can be used to define a rigging vector adapted to $\p\Si$, which is given by
\be
n = \f1{\sqrt 2}(e^\ba\t \mp e^{-\ba} \hat r).
\ee
Now $(l,n)$ and $(\t,\hat r)$ provide two possible basis for the time-like plane normal to $\p\Si$.
This change of basis is used to determine the corner terms required in the action by the variational principle.

\subsection{Affine coordinates}

The fact that a null hypersurface is ruled by null geodesics endows it with a preferred class of foliations, 
in which $\l$ is a parameter along the geodesics, $l^\m\p_\m \l=1$.
To use this parameter as one of the coordinates, we fix an initial cross-section of $\cN$, say at $\l=0$, define angular coordinates $x^A$ there and then Lie-drag them along $\cN$. 
This defines a coordinate system with vanishing shift vector, 
\be\label{affinemet}
l^a=(1,0,0), \qquad \pbi{g_{ab}}=\mat{0}{0}{}{\g_{AB}}.
\ee
These coordinates satisfy the  partial Bondi gauge. We have $l_\m=(g_{\l\Phi},0,0,0)$ and $g_{\Phi\l}=1/g^{\Phi\l}$, therefore this choice of tangent vector corresponds to the `canonical normalization' for $f$. This is an example of a situation in which $f$ is metric-dependent.
We can complete this partial gauge fixing on $\cN$ with a fourth condition, for instance redefining $\Phi$ so that $g_{\l\Phi}\eqonN -1$.
The metric now satisfies 
\be\label{nonaffinegf}
g_{\l\l}=O(\Phi), \qquad g_{\Phi\l}=-1+O(\Phi), \qquad g_{\l A}=O(\Phi),
\ee
and it is fully gauge-fixed on $\cN$. 

The coordinate system can be further specialized if we require the  parameter to be affine, namely
\be
\ella:=\f d{d\l}, \qquad \ella^\m\na_\m \ella^\n\eqonN0.
\ee
This condition fixes the first-order extension of the metric component $g_{\l\l}$ so that $\p_\Phi g_{\l\l}\eqonN 2\p_\l g_{\Phi\l}$.\footnote{
This follows from $\G^\m_{\l\l}\eqonN0$, which by invertibility of the metric is equivalent to 
$2 \p_\l g_{\m \l} - \p_\m g_{\l \l} \eqonN 0$. This is identically satisfied by \eqref{affinemet} for $\m=(\l,A)$, and thus reduces to the single equation given in the text.
}
Since one can always choose the adapted coordinate $\Phi$ 
such that $g_{\Phi\l}=-1+O(\Phi)$, in that gauge we have $g_{\l\l}=g^{\Phi\Phi}=O(\Phi^2)$. \footnote{Notice that this would be a `generalized' diffeomorphism, not invertible at the hypersurface, pretty much like going from static Schwarzschild coordinates to Eddington-Finkelstein is singular at the horizon.}
At this point,
\be\label{minimalgf}
g_{\l\l}=O(\Phi^2), \qquad g_{\Phi\l}=-1+O(\Phi), \qquad g_{\l A}=O(\Phi),
\ee
and the rest of the metric is arbitrary.
The condition of affinity can always be imposed via gauge-fixing, but we see that it is not a characteristic of the hypersurface coordinates alone, since involves the first-order extension of the metric. 

In the affine coordinate system, any normal vector in the equivalence class satisfies
\be\label{lfN}
l^\m \eqonN f \ella^\m
\ee
and 
\be\label{kf}
k = l^\m \p_\m \ln{f}.
\ee
Hence it is affine iff $f$ is chosen constant in $\l$, namely $\pounds_l f=0$. 
Furthermore since $\p_\Phi g^{\Phi\Phi}\eqonN0$, any extension of $l$ with $v\cdot l\eqonN 0$ satisfies 
$
\p_n l^2\eqonN0,
$
namely it is null at first-order off the hypersurface. We also recall that in affine coordinates $\bar k=0$.

This coordinate system can be extended to a neighbourhood of $\cN$ as follows (see e.g. \cite{Ashtekar:2021kqj}).
We shoot geodesics off $\cN$, and Lie drag $x^a$ along them.
Namely, we have
\be
n_o^\m = \f{\p}{\p\Phi},  \qquad n_o^\m\na_\m n_o^\n=k_{n_o}n_o^\n, \qquad \pounds_{n_o} x^a=0.
\ee
We can then completely fix the bulk coordinate gauge freedom if we require that $(i)$ $n_o$ is null everywhere, $(ii)$ $\Phi$ is affine (hence $k_{n_o}=0$), and $(iii)$
it is the gradient of the foliation of constant $\l$ on $\cN$, namely $n_o\eqonN- d \l$.
The last condition in particular means that $n_o$ gives a choice of rigging vector for $l_o$ adapted to the $\l$ foliation.
In terms of metric components, $(i)$ fixes $g_{\Phi\Phi}=0$, then $(ii)$ requires $\G^\m_{\Phi\Phi}=0$, which in turns implies $\p_\Phi g_{\Phi\m}=0$.
Finally, $(iii)$ fixes $g_{\Phi\m}\eqonN (-1,0,0,0)$.
The resulting coordinates $(\l,\Phi,x^A)$ are defined in a caustic-free open neighbourhood of $\cN$, in which the metric reads
\be
g_{\m\n} = \left(\begin{array}{ccc} 
\Phi^2 F & -1 & \Phi P_A \\  & 0 & 0 \\ & & \g_{AB}
\end{array}\right),
\qquad 
g^{\m\n} =\left(\begin{array}{ccc} 
0 & -1 & 0 \\  & -\Phi^2(F-P^2) & \Phi P^A \\ & & \g^{AB}
\end{array}\right),
\label{gabhay}
\ee
where $F, P_A$ and $\g_{AB}$ are arbitrary metric coefficients. 
With this gauge fixing, 
$\bar{\eta}_A = \G^{\Phi}_{A \Phi}  = - P_A/{2}$ and $\bar k=0$.
The latter makes it manifest that the extrinsic geometry captured by the inaffinity, or more precisely by $\bar k$, describes whether the hypersurface is described in affine coordinates or not, thus being faithful to its name.
We stress that what makes $\l$ an affine parameter on $\cN$ is not so much $g_{\l\Phi}=-1$ but $g_{\l\l}=O(\Phi^2)$.
This coordinate system can always be reached, and if one restricts the residual diffeomorphisms to preserve it, the whole extension of $\xi$ in the neighbourhood is fixed. 
On the other hand, if one relaxes it and requires only the minimal conditions \eqref{minimalgf}, only the first order extension $\hat \xi^\Phi$ is fixed, whereas $\hat\xi^\l$ and $\hat\xi^A$ remain arbitrary. 

We can now choose an extension of $l$ such that
\be\label{lf}
l^\m = f \ella^\m\,
\ee 
everywhere in the chart, and not only at $\cN$. This is achieved taking $v = \Phi Fd \l + P_Adx^A$. 
This extension is not hypersurface-orthogonal nor null nor geodesic, except at $\cN$.
But it satisfies $l\cdot v \eqonN 0$ and it is thus null at first order around $\cN$. 

Summarizing, affine coordinates on $\cN$ depend on extrinsic properties of the metric, and give us \eqref{minimalgf} and \eqref{kf}. The normal in these coordinates reads \eqref{lfN}, is in general not null at first order, but this can be easily achieved choosing for instance the extension \eqref{lf}.\footnote{Another convenient extension is $dl\eqonN 0$, which implies instead $v=-df+\Phi v'$, namely $l=-d(f\Phi)+\Phi^2v'$ is a gradient on $\cN$, and $v\cdot l\eqonN -\p_l f$.
Then choosing $\Phi'=f\Phi$ or more in general $f$ time independent is also enough to have $v\cdot l\eqonN0$ hence $\p_n l^2\eqonN 0$. 
So taking $dl\eqonN 0=\pounds_l f$ is another solution than \eqref{lf} to have $\p_n l^2\eqonN 0$.}

\section{Null symplectic potential}\label{SecNullTh}

We start from the standard Einstein-Hilbert symplectic potential 
\be\label{ThEH}
\th^{\sscr EH} =\f1{3!}\th^{\sscr EH}{}^\m \eps_{\m\n\r\s}~dx^\n\w dx^\r \w dx^\s , \qquad
\th^{\sscr EH}{}^\m = 2 g^{\r[\s} \d \G^{\m]}_{\r\s}, 
\ee
and consider the most general expression for its
pull-back on a null hypersurface. This was computed for instance in \cite{Oliveri:2019gvm},\footnote{With $\om$ here defined with opposite sign, as to match \cite{Ashtekar:2000hw,Ashtekar:2004cn}. 
We took the time to accurately translate notations to prove that it is indeed  equivalent to the one computed in \cite{Parattu:2015gga}, including the corner term, thus answering the question left open in \cite{Oliveri:2019gvm}. In doing so we realized that the statement in \cite{Oliveri:2019gvm} that taking an extension with $\p_n l^2\neq 0$ produces an extra term $\p_n l^2 n^\m \d l_\m$ in $\th^{\sscr EH}$ was incorrect. We thank Laurent Freidel for pointing this out to us. The final expression generalizes the one of \cite{Lehner:2016vdi} which assumes $\d l^\m=0$, and the one of \cite{Hopfmuller:2016scf} which assumes $\d l_\m = -n^\rho l^\sigma\d g_{\rho\sigma} l_\m$, the latter implying that $n_\m\d l^\m=0$.
It also generalizes the one of \cite{Chandrasekaran:2020wwn} -- contrarily to what there stated --, which assumes $\d l_\m=0$, see \eqref{ThNPi}. We will come back to these restrictions and their motivations below. We hope that no confusion arises because the same letter $\th$ appears as both the symplectic potential integrand and the expansion of $l$. The risk should be reduced by the fact that the letter used for the symplectic potential always comes with labels such as $^{\sscr EH}$ or $'$.
}
and reads
\begin{equation}\label{ThN1}
\pbi{\th}^{\sscr EH}=\big[ (\s^{\m\n}+\f\th2\g^{\m\n}) \d \g_{\m\n} - 2  \om_{\m}\d l^\m + 2 \d \left(\th + k \right) \big]\eps_\cN
+ d\vartheta^{\sscr EH}.
\end{equation}

This expression holds for arbitrary variations on the null-hypersurface: the only restriction made is to preserve the null nature of the hypersurface, namely
\be\label{nullpreserving}
l_\m\d l^\m\eqonN0. 
\ee
In particular, it is valid for a field-dependent $f$, so that 
\be
\d l_\m \eqonN \d\ln f \, l_\m
\ee
doesn't need to vanish. If $f$ is field-independent, 
$\d l_\m\eqonN0$ or equivalently $n^\m\d l_\m\eqonN0$.

The expression \eqref{ThN1} does not depend on rescalings of $l$ nor on the choice of auxiliary vector $n$. Independence from rescalings follows from its invariance under the class-III spin-boost transformations \eqref{classIII}, and implies  independence from changes of embeddings $\Phi\to \Phi'(\Phi)$.
Independence from $n$ follows from the invariance under class-I Lorentz transformations \eqref{classI}. See App.~\ref{AppLorentz} for proofs. 
The invariance is only a property of the full expression. The individual quantities are not, as summarized in 
Table~\ref{invTable}. This is relevant for various considerations that we do below.

The variation of the inaffinity can be written in terms of $\d  l_\m$ and $\d l^\m$:
\begin{align}
\d k &=k n^\mu \delta l_\mu - 2n^\mu  \nabla_{(\mu}l_{\nu)} \delta l^\nu-  2 n^\mu l^\nu \nabla_{(\mu} \delta l_{\nu)}  + \frac{1}{2} n^\mu \nabla_\mu \delta l^2 \nn\\ \label{dkgen}
&= ( k n^\m +\f12 n^{\n}\na_\n l^{\m} - \f12 l^{\m}n^{\n}\na_\n - n^{\m} l^{\n}\na_\n )\d l_\m 
- (n^\n\na_{\m}l_{\n}+\f12 n^\n\na_\n l_\m-\f12 l_\m n^\n \na_\n )\d l^\m. 
\end{align}
Notice that presence of normal derivatives on the variations, which imply that $\d k$ varies even if we fix both $l_\m$ and $l^\m$ on the hypersurface:
\be \label{dkspecial}
\d l^\m= \d l_\m \eqon\cN 0 \qquad \Rightarrow \qquad 
\d k = - \f12 n^\m\na_\m (l_\n l_\r \d g^{\n\r}).
\ee
This vanishes if we restrict the variations to preserve affine coordinates. Therefore $\d k$ is an independent variation because it captures the possibility of varying the metric between the form \eqref{nonaffinegf} and \eqref{minimalgf}, which are both consistent with fixing $l_\m$ and $l^\m$ on the hypersurface.

Finally, the corner term is  \cite{Hopfmuller:2016scf,Oliveri:2019gvm}
\be\label{vthEH}
\vth^{\sscr EH}:=n^\m\d l_\m \eps_S - i_{\d l}\eps_\cN =
\left(n^\m\d l_\m + n_\m\d l^\m \right)\epsilon_S - n\w i_{\d l}\eps_S.
\end{equation}
The last term vanishes if we pull-back on a space-like cross section with $n$ adapted to it.

\subsection{Phase space polarizations}

We would like to manipulate the RHS of \eqref{ThN1} so to put it in the form
\be\label{th'}
\pbi{\th}=\th'-\d\ell+d\vth,
\ee
where $\th={\th}^{\sscr EH}$, and $\th'=p\d q$ for a given choice of polarization of the (kinematical) phase space on the boundary.\footnote{This notion of polarization should be used only at the level of the kinematical phase space. In fact, on a null-hypersurface the momenta do not depend on velocities, hence there are second class constraints and the actual symplectic structure is given by a Dirac bracket and not the initial, kinematical Poisson bracket, see e.g. \cite{Torre:1985rw,Alexandrov:2014rta}.}
This form will be useful to discuss two different but related contexts: the variational principle, and the definition of charges using covariant phase space methods.

If we attempt to interpret \eqref{ThN1} as a symplectic potential in $p\d q$  form, we see that the $q$'s appearing are not independent,
since $\g_{\m\n}$ also determines $\th$ through \eqref{defB}. To put it in diagonal form, we need two steps. 
The first is to observe that the variation of the volume form  has  two components, 
\be\label{trecarte}
\d\eps_\cN =\left( \f12 \g^{\m\n}\d \g_{\m\n} + n_\m\d l^\m \right) \eps_\cN,
\ee
as can be established starting from the identity
\be\label{detidentity}
\f12 g^{\m\n}\d g_{\m\n} = \f12 \g^{\m\n}\d \g_{\m\n} + n_\m\d l^\m- n^\m\d l_\m.
\ee

Using \eqref{trecarte}, the pull-back \eqref{ThN1} can be rewritten in the form
\begin{align}\label{ThN2}
\pbi{\th}^{\sscr EH} = \big[ \s^{\m\n} \d \g_{\m\n}  + \pi_\m \d l^\m +2\d(\th+k)
\big]\eps_\cN +\th\d\eps_\cN+ d\vartheta^{\sscr EH}, 
\end{align}
where
\be
\pi_\m:=-2\left(\om_\m+\f\th2n_\m\right) = 2 \left(\eta_{\m}+\left(k-\f\th2\right)n_\m\right).
\ee
The second step is to integrate by parts in field space the third term. This leads to
\begin{align}\label{Thdiag}
\pbi{\th}^{\sscr EH} =  \th^{\sscr D}
 - \d\ell^{\sscr D} 
+ d\vartheta^{\sscr EH},
\end{align}
where
\be\label{ThD}
\th^{\sscr D}:=\s^{\m\n} \d \g_{\m\n} \eps_\cN + \pi_\m\d l^\m\eps_\cN -(\th+2k)\d\eps_\cN,
\ee
and
\be\label{ellD}
\ell^{\sscr D} := - 2W \eps_\cN = -2(\th+k)\eps_\cN = -2k\eps_\cN -2d\eps_S.
\ee
The last equality follows from \eqref{des}, and can be used to simplify the boundary term reabsorbing its dependence on the expansion in the corner term, and work with
\be\label{ellD'}
\ell^{\sscr D}{}' = -2k\eps_\cN, \qquad \vth^{\sscr EH}{}'=\vth^{\sscr EH} +2\d\eps_S.
\ee

We can now identify $\th'=\th^{\sscr D}$,  which is diagonal form $p\d q$ with 
$\smash{q=(
\pbg,l^\m)}$.
The $q$ terms only involve the intrinsic geometry, therefore the symplectic potential is in the form of a Dirichlet polarization,
whence the D label. The diagonalization obtained involves the sum of three pairs of configuration variables and momenta that can be characterized as spin 2, 1 and 0, as discussed for instance in \cite{Hopfmuller:2016scf}. 
Denoting $\varepsilon:={\sqrt{-g}}/f$, we can write the conjugate momenta as densities,
\be
\tl\pi^{\m\n}:= \varepsilon\s^{\m\n}, 
\qquad \tl\pi_\m:=2 \varepsilon \left(\eta_{\m}+\left(k-\f\th2\right)n_\m\right), \qquad \tl\pi:=-{\varepsilon} \left(\th + 2k \right).
\ee

The first term in \eqref{ThD} can be equally written as
\be\label{spin2switch}
\s^{\m\n}\d\g_{\m\n} \, \eps_\cN= -\s_{\m\n}\d(\g^{\m\n}\,\eps_\cN)
=-\g_{\m\n}\d \s^{\m\n}\eps_\cN, 
\ee
where $\g^{\m\n}\,\eps_\cN$ is manifestly conformal invariant. In other words, only the trace-less part of the metric perturbations enters here.
This is the spin-2 pair. 

The spin-1 pair has three components, two `transverse' ones whose momentum is the twist, and a `longitudinal' one proportional to $n_\m \d l^\m$.
The longitudinal variation is there to compensate the dependence of $\g^{\m\n}\d\g_{\m\n}$ on the choice of rigging vector. This dependence 
prevents the interpretation of $\g^{\m\n}\d\g_{\m\n}$ as the variation of the volume form, 
see \eqref{trecarte}, and can be removed if we restrict the variations to satisfy
\be\label{ndl0}
n_\m\d l^\m \eqonN 0 \qquad \Rightarrow \qquad  
 \f12 \g^{\m\n}\d \g_{\m\n} \eps_\cN \eqonN \d\eps_\cN.
\ee
This restriction can be achieved in two ways. The first is to choose the metric-dependence of $f$ such that $\d l_\m\eqonN-n^\rho l^\sigma\d g_{\rho\sigma} l_\m$ \cite{Hopfmuller:2016scf}, from which \eqref{ndl0} follows. The second is to  choose first a foliation, and then the `canonical' normalization for $f$ \cite{Oliveri:2019gvm}.

The split into spin pairs is appealing, but it is not canonical: the spin-2 pair is class-III invariant but not class-I invariant, and the spin-1 and spin-0 pairs mix up under class I and class III transformations, as can be easily seen using the formulas given in App.~\ref{AppLorentz}. 
In other words, the spin pairs are not singled out by the geometry of the hypersurface, but require additional non-geometric choices of normal representative and of rigging.\footnote{They are thus singled out by a rigged Carollian structure, see previous footnote~\ref{FootCarroll}.}

There is however a more serious problem with $\th^{\sscr D}$: it is not class-III invariant even as a whole.
This can be checked explicitly, but it can be more easily read off the fact that the standard symplectic potential is class-III invariant, whereas
the boundary Lagrangian and corner potential are not:
\begin{align}
\d\ell^{\sscr D}\to \d\ell^{\sscr D}-2\d(\pounds_{l}\ln A\,\eps_\cN), \qquad \vth^{\sscr EH}\to \vth^{\sscr EH}-2\d\ln A\,\eps_S.
\end{align}
Adding up and using \eqref{anvedi}, we find
\be
\th^{\sscr D}\to\th^{\sscr D} -2(\pounds_{\d l}\ln A-\th\d\ln A)\eps_\cN - 2\pounds_l\ln A\,\d\eps_\cN.
\ee
We will see below how this issue affects both the variational principle and the covariant phase space.

Notice that insofar as the rescaling is field-independent, $\d A=0$ and the non-invariance of $\th^{\sscr D}$ stems from the boundary Lagrangian alone. Let us first restrict attention to this case.
One way to obtain class-III invariance is to restrict the variations so that $\d l^\m$ and $\d\eps_{\cN}$ vanish. 
The first is possible without loss of generality by a restriction on the coordinates, whereas the second is a strong restriction on the dynamics, and we are interested in avoiding it.

One way to do so is to impose  $\d k=0$, as was done in the CFP paper \cite{Chandrasekaran:2018aop}.
If we do so, then
\be
\pbi{\th}^{\sscr EH}|_{\d k=0} = \big[ \s^{\m\n} \d \g_{\m\n}  + \pi_\m \d l^\m +2\d\th\big]\eps_\cN +\th\d\eps_\cN+ d\vartheta^{\sscr EH}
=\th^{\sscr CFP}-\d\ell^{\sscr CFP}+ d\vartheta^{\sscr EH}, 
\ee
where
\be\label{ThDc}
\th^{\sscr CFP}:=\big(\s^{\m\n} \d \g_{\m\n}  + \pi_\m \d l^\m\big)\eps_\cN - \th\d\eps_\cN, \qquad \ell^{\sscr CFP}:=-2\th\eps_\cN.
\ee
This choice makes the boundary Lagrangian and the symplectic potential class-III invariant. 

If we impose instead $\d l_\m=0$, we can replace $k$ with $\bar k$ in \eqref{ThN2} and obtain
\be
\pbi{\th}^{\sscr EH}|_{\d l_\m=0} =\th^{\sscr \bar D}-\d\ell^{\sscr \bar D}+ d\vartheta^{\sscr EH}, 
\ee
where
\be\label{ThbarD}
\th^{\sscr \bar D}:=\s^{\m\n} \d \g_{\m\n} \eps_\cN + (\pi_\m+2\p_\m\ln f) \d l^\m\eps_\cN -(\th+2\bar k)\d\eps_\cN,
\qquad \ell^{\sscr \bar D}:=-2(\th+\bar k)\eps_\cN.
\ee
This boundary Lagrangian is independent of $f$, but not of reparametrizations of $\Phi$. This is  only a partial resolution of the ambiguities, and 
class-III invariance is not achieved.

Remarkably, there is a solution
that does not require any restriction on the variations,
but instead performing an integration by parts in field space of the spin-0 term:
\be\label{spin0byparts}
-(\th+2k)\d\eps_\cN = -\d[(\th+2k)\eps_\cN] + \d(\th+2k)\eps_\cN.
\ee
This leads to
\begin{align}\label{ThYdec}
\pbi{\th}^{\sscr EH} =  
\th^{\sscr Conf} - \d\ell^{\sscr Conf}
+ d\vartheta^{\sscr EH},
\end{align}
where now
\be\label{ThY}
\th^{\sscr Conf}:=[\s^{\m\n} \d \g_{\m\n}  + \pi_\m\d l^\m +\d(\th+2k)]\eps_\cN,
\ee
and
\be\label{ellY}
\ell^{\sscr Conf}:= - \th\eps_\cN = -d\eps_S.
\ee
The new boundary Lagrangian and symplectic potential are class-III invariant.
This shows the type of valuable insights that can be obtained using the freedom of changing potential via \eqref{th'}.

The decomposition \eqref{ThY} corresponds to a change of polarization in the phase space that identifies as configuration variables the conformal class of the 2d metric -- equivalently the shear, recall \eqref{spin2switch} and discuss around there --, the tangent vector, and the spin-0 momentum $\th+2k$ instead of the volume form. 
We will show that it leads to a Noether flux-balance law with no anomaly term.
This is related to that fact that this choice makes the boundary Lagrangian unambiguous even without any restriction on the variations. 
Conformal boundary conditions appear thus to be better behaved, something argued for also in the time-like case \cite{Anderson:2006lqb}.
This polarization was referred to as $\th^{\sscr Y}$ in \cite{Rignon-Bret:2023fjq}, by analogy with York's conformal boundary conditions for the time-like case.
A similar decomposition to \eqref{ThY} was considered also in \cite{Hopfmuller:2018fni}, while looking for thermodynamical interpretations of the symplectic potential.  However they included the additional restriction $n_\m \d l^\m=0$, which spoils class-I invariance.

There are other integrations by parts in field space that could be considered.
One could change the spin-0 sector with different numerical factors. For instance the investigations of black hole entropy by Chandrasekaran and Speranza (CS) in \cite{Chandrasekaran:2020wwn}
motivates the choice
\be
\ell^{\sscr CS} = -(k+2\th)\eps_\cN = \ell^{\sscr D} +k\eps_\cN.
\ee
The motivation from black hole entropy will be briefly explained below, but notice that this choice is not class-III invariant.
On the other hand, a change of polarization in the spin-1 pair seems of little use, since we cannot treat $\eta_\m$ as an independent configuration variable from the spin-2 pair. This is due to the fact that the constraint equations on $\cN$ relates $\eta_\m$ 
to the radiative data which are contained in the spin-2 pair.
For completeness, we report in App.~\ref{AppPolar} an exploration of alternative polarizations and their boundary conditions. 

In the rest of the paper we will study 
how changing the boundary Lagrangian as in the examples above affects the variational principle and 
the construction of gravitational charges. 
With the exception of \eqref{ThbarD}, the boundary Lagrangians that we consider have the same functional dependence, and differ only by numerical factors. 
This is similar to what we had in the case of a time-like boundary \cite{Odak:2021axr}, and allows us to treat all cases at once
writing the boundary Lagrangian in parametric form as
\be\label{ellfamily}
\ell^{(b,c)}=-(bk +c\th)\eps_\cN.
\ee
A boundary Lagrangian of this family is class-III invariant for $b=0$ and any value of $c$.
The specific examples described above correspond to: 
\be
\begin{array}{c|cccc} & {\rm D} & {\rm CFP} & {\rm Conf} & {\rm CS} \\ \hline
b & 2 & 0 & 0 & 1 \\
c & 2 & 2 & 1 & 2
\end{array}
\ee
The option \eqref{ThbarD} could be included adding a third parameter, but we have seen that it is only a partial solution and will thus be considered less in the following.
The symplectic potential corresponding to this family is
\begin{align}\label{Thbc}
\th^{(b,c)} = \big[ \s^{\m\n} \d \g_{\m\n}  + \pi_\m \d l^\m +(2-b)\d k+(2-c)\d\th \big]\eps_\cN 
-(bk+(c-1)\th)\d\eps_\cN,
\end{align}
and we repeat for convenience of comparison the four particular cases discussed earlier:
\begin{subequations}\begin{align}
& \th^{\sscr D}:=(\s^{\m\n} \d \g_{\m\n}  + \pi_\m\d l^\m)\eps_\cN - (\th+2k)\d\eps_\cN, \\
& \th^{\sscr CFPk}:=\big(\s^{\m\n} \d \g_{\m\n}  + \pi_\m \d l^\m +2\d k\big)\eps_\cN - \th\d\eps_\cN, \\
& \th^{\sscr Conf}:=[\s^{\m\n} \d \g_{\m\n}  + \pi_\m\d l^\m +\d(\th+2k)]\eps_\cN, \\
& \th^{\sscr CS}:=\big(\s^{\m\n} \d \g_{\m\n}  + \pi_\m \d l^\m +\d k\big)\eps_\cN -(k+ \th)\d\eps_\cN.
\end{align}\end{subequations}
Here CFPk stands for the extension of the CFP case to $\d k\neq 0$.
Although both $\th^{\sscr CFPk}$ and $\th^{\sscr Conf}$ are III-invariant, only the latter is in diagonal form for the general case with $\d k\neq 0$, since in the former the volume form appears both as $q$ and as $p$. More in general, any potential with $b=0$ is III-invariant, 
but only the one with $c=1$ is in diagonal form. For $\d k=0$, both $\th^{\sscr CFP}$ and $\th^{\sscr Conf}$ are diagonal and class III-invariant for $\d A=0$.

So far we have assumed that $\d A=0$. This condition is satisfied if we restrict the class of allowed normals to satisfy $\d l^\m\eqonN 0$ or $\d l_\m\eqonN 0$.
If one relaxes both conditions and allows for $\d A\neq 0$, then a class-III invariant boundary Lagrangian is no longer sufficient to have a class-III invariant symplectic potential, because of the contribution from $\vth^{\sscr EH}$. A possibility would then be to absorb $d\vth^{\sscr EH}$ in the definition of $\th'$. This is possible of course, however it would spoil the idea that $\th'$ should be in diagonal $p\d q$ form. Furthermore, we will see that there are other reasons to impose 
$\d l^\m\eqonN 0$ or $\d l_\m\eqonN 0$, as well as to keep $\vth^{\sscr EH}$ out of $\th'$. For this reason we keep $\vth^{\sscr EH}$ in the corner term.

\section{Conservative boundary conditions and the variational principle}

In the study of the variational principle, one wants to find suitable boundary conditions that make the variation of the action vanish everywhere on-shell, including at the boundary. 
In this context, \eqref{th'} is useful to identify the required boundary and corner terms to be added to the action for the allowed boundary conditions.
Suppose that we find a decomposition like \eqref{th'} with a certain $\th'=p\d q$, and such that adding $\vth$ to the contribution coming from the part of the boundary complementary to $\cN$ we get a total variation, call it $\d c$. 
We can then conclude that the boundary conditions identified by $\d q=0$ provide a well-defined variational principle,
once the action is supplemented with the boundary term $\ell$ as well as the corner term $c$.

In this Section we show how the different polarizations of the null symplectic potential give a variational principle with different boundary conditions. We first review two known but non-trivial facts about Dirichlet boundary conditions, namely that one has to fix one more condition than the intrinsic geometry \cite{Parattu:2015gga}, and that the resulting boundary terms are ambiguous \cite{Lehner:2016vdi}. We then show that the alternative conformal boundary conditions improve this problem.

\subsection{Dirichlet boundary conditions and their ambiguity}

Dirichlet boundary conditions hold fixed the intrinsic geometry. In the case of a null hypersurface, we could take this to mean
\be\label{Dbc}
\g^{\m\r}\g^{\n\s}\d\g_{\r\s}\eqonN 0, \qquad \d l^\m\eqonN 0.
\ee
The first condition is class-III invariant, but the second only if $\d A=0$. Nonetheless 
they imply $\d\eps_\cN\eqonN0$ thanks to \eqref{trecarte}, therefore they fix entirely the intrinsic geometry. 
On-shell of these conditions 
\eqref{ThN1} gives
\begin{equation}\label{ThDon}
\pbi{\th}^{\sscr EH}= -\d\ell^{\sscr D}  + d\vth^{\sscr EH},
\end{equation}
with $\vth^{\sscr EH}=n^\m \d l_\m \eps_S$. 
The first term on the RHS  is a total variation, and can be eliminated if the boundary Lagrangian
\eqref{ellD} is added to the initial action.  This is the equivalent of the Gibbons-Hawking-York term, and can even be written in exactly the same form as the divergence of the normal using \eqref{Wtrace}. 
Notice also that \eqref{Dbc} also imply $\d\th\eqonN 0$, hence the only relevant term in \eqref{ellD} is the inaffinity $k$. 
It is then equivalent to work with this boundary Lagrangian or the alternative choice \eqref{ellD'}.

The second term on the RHS is not a total variation, but it can be shown that once it is added to the contribution coming from the rest of the boundary, one obtains a total variation \cite{Hartle:1981cf,Hayward:1993my,Lehner:2016vdi,Jubb:2016qzt,Oliveri:2019gvm}.
For instance if the null boundary is joined to a space-like boundary $\Si$, 
\be\label{corner}
\vth^{\sscr EH} + \vth^{\sscr EH}_{\Si} =-2 \d\ba \epsilon_S,
\ee
where $\ba$ is defined by \eqref{ldottau}.
This is a total variation under Dirichlet boundary conditions, since the first of \eqref{Dbc} implies $\d\eps_S\eqon\cN0$. It can thus be compensated by the corner  Lagrangian
\be\label{ellH}
l^{\sscr H} :=2\ba \eps_S.
\ee 
Here H stands for Hayward.
See \cite{Hayward:1993my,Lehner:2016vdi} for other examples of joints and their corner terms.

We see that the Dirichlet variational principle is not well-defined for the Einstein-Hilbert action with a null boundary, and one needs to supplement the action with the boundary terms $\ell^{\sscr D}$ and $\ell^{\sscr H}$, in analogy with what happens with other types of boundaries. 
This is the result that one typically finds in the literature \cite{Parattu:2015gga,Lehner:2016vdi,Jubb:2016qzt}.
The problem that was raised in \cite{Lehner:2016vdi} is that these boundary and corner terms are ambiguous and non-geometric: they involve quantities like $k$ and $\ba$ which depend on the choice of normal representative and on changes of embedding. 
In our language, they are not class-III invariant.
A solution proposed in \cite{Lehner:2016vdi} was to add the non-local boundary term $(\th\ln\th)\eps_\cN$.
An alternative solution that doesn't involve the non-local counterterm would be to 
work with class-III invariant quantities only.

As we have seen at the end of the previous Section, this can be achieved in various ways. 
One is to add the condition $\d k\eqonN 0$.
The  boundary Lagrangian for this variational principle is the class-III invariant choice $\ell^{\sscr CFP}$ and it is unambiguous.
In fact, it is spacetime exact, hence it can be reabsorbed in the corner term, leading to a variational principle without any boundary term along the null hypersurface. 
The problem raised in \cite{Lehner:2016vdi} is however not completely solved, because there is still the need for corner terms like \eqref{ellH}, which maintain their ambiguity. 

Next, let us consider the condition $\d l_\m \eqonN 0$. Adding it is quite natural, since the combination $\d l^\m=\d l_\m \eqonN 0$ is equivalent to $l_\m \d g^{\m\n}\eqonN 0$ and as such, manifestly class III-invariant. 
A boundary Lagrangian for the Dirichlet variational principle supplemented by $\d l_\m\eqonN 0$ is given in \eqref{ThbarD}, but as remarked there, 
this removes only the ambiguity under change of $f$ at fixed embedding, and not under changes of embedding.
This would then only be a partial resolution to the problem of ambiguities. The reason why we have class-III invariant boundary conditions but fail to have a fully class-III invariant boundary Lagrangian is that $\ell^{\bar D}$ transforms under class-III with a inhomogeneous term proportional to $\pounds_l \ln A$, and its variation is zero under the above conditions. 

As for the corner terms, the additional condition means that
\eqref{vthEH} vanishes. This does not remove the Hayward corner term \eqref{ellH} from the variational principle, since it is still needed to cancel a contribution to the variation coming from the space-like boundary. But it removes it in the case of a corner between two null boundaries with the same boundary conditions. In the latter case there is no corner ambiguity. 

It follows that if we take both additional restrictions,
\be\label{DbcS}
\g^{\m\r}\g^{\n\s}\d\g_{\r\s}=\d l^\m=\d l_\m=\d k \eqonN 0,
\ee
 this strengthened Dirichlet variational principle is well-defined with no contribution from the null boundary, and all potential ambiguities reduced to a choice of normal at the corner between a null and a non-null boundary. 
 
The importance of choosing the right boundary term goes beyond the variational principle. In \cite{Lehner:2016vdi}, it was discussed its relevance to in the context of the `action=complexity' proposal in AdS/CFT holography. Below, we will see how it affects the charges constructed with covariant phase space methods.
In this application, only the ambiguity of the boundary Lagrangian matters, and not the one of the corner terms used in the variational principle.

\subsection{Conformal boundary conditions}

Consider now the alternative polarization \eqref{ThY}. This vanishes for the conformal boundary conditions
\be\label{Ybc}
\d\s^{\m\n}\eqonN0, \qquad \d l^\m\eqonN 0, \qquad \d(\th+2k)\eqonN 0.
\ee
The first two are equivalent to Dirichlet except for $\d \eps_\cN\eqonN 0$, which is replaced by the third one above. Since $\th$ is intrinsic and $k$ depends on transversal derivatives of the metric, the last condition is of Robin type. 
With respect to the general family \eqref{Thbc}, 
$\th^{\sscr Conf}$ corresponds to $b=0$ and $c=1$, and it is easy to see that it is the only possibility that would allow conservative boundary conditions with $\d\eps_\cN\neq 0$.
We then have
\begin{equation}\label{ThYon}
\pbi{\th}^{\sscr EH}= -\d\ell^{\sscr Conf} 
+ d\vth^{\sscr EH}.
\end{equation}

The interesting remark is that this boundary Lagrangian is  class-III invariant, hence geometric and not ambiguous. 
Changing polarization resolves the problem of ambiguity of the boundary Lagrangian without adding counter-terms, nor any of the additional restrictions $\d l_\m=0$ or $\d k=0$ considered above. 
If we do include the $\d k=0$ restriction, then the conformal and Dirichlet polarization boil down to the same boundary conditions, and their boundary Lagrangians 
 \eqref{ellY} and \eqref{ThDc} are indeed the same up to corner terms.

Next, we look at the corner terms.
$\vth^{\sscr EH}$ is the same as before, therefore we still have \eqref{corner} when looking at the joint between a null and a space-like boundary. This is no longer a total variation, because  \eqref{Ybc} does not imply $\d\eps_S\eqonN0$. Therefore, to have a well-defined variational principle we need to add the corner boundary condition
\be\label{cYbc}
\d\eps_S\stackrel{S}=0.
\ee
The need for an additional condition on top of \eqref{Ybc} seems reasonable because the expansion is the derivative of the 2d metric, hence one is missing an initial datum when providing boundary data in terms of the expansion.
Upon doing so, the required action corner terms in the variational principle are the same as in the Dirichlet case, e.g. \eqref{ellH}. Therefore even though these boundary conditions eliminate the ambiguity of the null boundary Lagrangian, they do not eliminate the ambiguity of the corner terms, at least in so far as they are completed with \eqref{cYbc}. If we further add $\d l_\m=0$ then as before we remove the need of corner terms between null boundaries.

The boundary Lagrangian $\ell^{\sscr Conf}$ is spacetime exact, hence it could also be absorbed into a modified $\vth$. In this case the conformal boundary conditions require no boundary Lagrangian at all. This however does not change the ambiguity of the corner terms, since one is adding a non-ambiguous term to the existing ambiguous one.

Having found a polarization with a class-III invariant boundary Lagrangian will be very useful for the construction of charges in the covariant phase space, which is what we turn to next. In that context, having corner terms in the action principle which are ambiguous is not important, because these don't enter neither the expression for the Noether current nor that for the Hamiltonian generator.
However from the point of view of the variational principle one may be interested in going further, and see whether it exists a completely unambiguous variational principle including the corner terms. A possibility would be to replace  \eqref{cYbc} with its conjugated corner variable, namely set $\d\ba\stackrel{S}= 0$.
No corner term in the action would then be needed. To that end, one should first study whether $\d(\th+2k)\stackrel{S}=0$ has any bearing on $\ba$.
We leave further investigations of this idea for future work.

\section{Leaky boundary conditions and covariant phase space}\label{SecLeaky}

Sachs' identification of constraint-free data on a null hypersurface \cite{Sachs:1962zzb} can be used to cut a distinction between physical and gauge degrees of freedom, and in turn understand which part of the conservative boundary conditions can be relaxed in order to allow flux of physical degrees of freedom through the boundary. Such leaky boundary conditions are useful in order to construct a covariant phase space that describes the evolution of dynamical gravitational systems using the flux defined by the symplectic potential.
As we have seen, the symplectic potential depends on the intrinsic and extrinsic geometry of the hypersurface, 
as well as on non-geometric quantities such as the choice of extension of the normal. Furthermore any split like \eqref{th'} can introduce a dependence of the individual terms on the scaling of the normal and on the choice of rigging vector. This dependence shows up in the covariant phase space in the form of anomalous transformations of the fields, or anomalies for short. The exact nature of the anomalies depends on the leaky boundary conditions chosen.
Different types have been explored in the literature, 
leading to different residual gauge transformations and thus different boundary symmetry groups. 
In this Section we characterize the different symmetry groups and compute their anomalies.
In the following Section we will see how the anomalies enter the gravitational flux and the Noether charges, and study how the flux-balance laws are reorganized when changing polarization of the symplectic potential.

Let us first talk about the variations that enter the symplectic potential.
The shear and expansion are determined by the induced metric, and the twist is also expected to be determined by the induced metric on-shell of the Einstein's equations. Therefore, the symplectic potential contains substantially only four independent variations:
$\d\g_{\m\n}$, $\d l^\m$, $\d l_\m$ and $\d k$.
The radiative data are contained in the first variation, so that one should definitely be left free in leaky boundary conditions allowing for gravitational flux. The question is then what to do with the remaining three. 

On the  hypersurface, we can always choose coordinates such that the tangent vector $l^\m$ has the simple form \eqref{affinemet}. Therefore $\d l^\m\eqonN 0$
can be interpreted as a restriction of the phase space to variations preserving this choice of coordinates. Furthermore, $l_\m$ is now metric dependent, unless we fix the $\Phi$ coordinate to have $g_{\Phi\l}=-1$, or equivalently we take the `canonical' normalization. With the first option, $\d l_\m\eqonN 0$ is also a restriction of the phase space preserving a certain choice of coordinates. Both coordinate choices are always achievable and don't restrict the physics of the system: they can be taken as part of the universal structure. Therefore it seems reasonable to impose  both restrictions,
and this is indeed the conclusion reached through the careful analysis done in \cite{Chandrasekaran:2018aop}.

The situation with $\d k$ is a bit more subtle, because one may expect that having eliminated the variations coming from $f$, this is a genuine variation of the extrinsic geometry that contains physics. 
But as we see from \eqref{dkspecial}, this variation captures the transversal derivative of $l_\m l_\n \d g^{\m\n}$, whose vanishing means that the coordinates are affine. 
Hence restricting $\d k$ to vanish or not means that the symmetry group of the covariant phase space preserves or not affine coordinates, namely metrics in the form \eqref{minimalgf} as opposed to \eqref{nonaffinegf}. And this seems merely a gauge statement.

This question was also left partially open in \cite{Chandrasekaran:2018aop}. In the main body of the paper $\d k$ is fixed to vanish, on the account that any two metrics with a  null hypersurface $\cN$ can be made to have the same pair $(l^\m,k)$ via a diffeomorphism on $\cN$, suggesting that $k$ should be taken as part of the universal background structure. Vanishing $\d k$ was also used in order to complete the Wald-Zoupas procedure and identify a non-ambiguous notion of charges. 
On the other hand, it was pointed out that the symplectic 2-form is not degenerate along any of the boundary diffeomorphisms. If one takes zero-modes of the symplectic 2-form with boundary terms included to be a definition of gauge, then every diffeomorphism of the boundary should be considered as a physical transformation.
This motivates the investigation of an enlarged phase space in which $k$ is allowed to vary, and possibly even $l^\m$.
As we will see, enlarging the phase space affects the symmetry groups and the associated transformations in field space, in particular their anomalies.

\subsection{Anomalies and class-III invariance}

The variations studied in Section~\ref{SecNullTh} 
keep the boundary fixed and null: $\d\Phi=0$ and $l_\m\d l^\m\eqonN 0$, or $\d g^{\Phi\Phi}\eqonN 0$ in adapted coordinates.
This means that $\Phi$ and $g^{\Phi\Phi}|_\cN$ are fixed background structures, whereas the rest of the metric can be varied freely. 
The split between dynamical (namely varying) and background (namely fixed) structures introduces a delicate aspect in the construction of the phase space, that of anomalies, which we now review. We use $\phi$ to denote a generic collection of dynamical fields, which we will specialize below to the spacetime metric.

We use the notation of \cite{Freidel:2021cjp} for the exterior calculus in field space. 
In particular, $\d$ is the exterior derivative in field space, $I_{X}$ the inner product with a vector field
\be
\hat X = \int X(\phi)\f{\d}{\d \phi},
\ee
and $\d_{X}=\d I_{X} + I_{X} \d $ is the Lie derivative.  Recall to avoid any confusion that the corresponding quantities for the exterior calculus in spacetime are denoted $d$, $i_\xi$ and $\pounds_\xi=d i_\xi +i_\xi d$.
We will often consider vector fields in field space whose components are spacetime diffeomorphisms, and which we denote by
\be
\hat \xi := \int \pounds_\xi \phi\f{\d}{\d \phi}.
\ee
It follows from this definition that the field-space Lie derivative of a dynamical field $\phi$ coincides with the spacetime Lie derivative, namely $\d_\xi\phi=\hat\xi(\phi)=\pounds_\xi\phi$, whereas for a background field $\chi$ we have $\d_\xi\chi=\hat\xi(\chi)=0$. 
If $\chi$ is not left invariant by the diffeomorphisms under considerations, $\pounds_\xi \chi\neq 0$ and therefore $\d_\xi$ and $\pounds_\xi$ have a different action,
and the field $\chi$ is thus non-covariant.
More in general, we say that a functional $F(\phi,\chi)$ in field space is covariant if the Lie derivatives coincide, $\d_\xi F=\pounds_\xi F$. This property is trivial for any functional that depends on the dynamical fields only, but may fail for functionals that depend on background fields as well. The difference $\d_\xi-\pounds_\xi$ then measures the non-covariance.

It is also important to introduce the anomaly operator
\be\label{Dxidef}
\D_\xi := \d_\xi-\pounds_\xi-I_{\d\xi},
\ee
which coincides with the non-covariance for field-independent diffeomorphisms and for field-space scalar functionals.
The third term in \eqref{Dxidef} is  relevant when acting on functionals of the fields that are forms in field space, as for example on the symplectic potential:
For a 1-form $F(\phi,\chi)\d\phi$ we have
\be
\d_\xi (F\d\phi) 
= \p_\phi F\d_\xi \phi \d\phi +F\d\d_\xi\phi = \p_\phi F\pounds_\xi \phi \d\phi +F\d\pounds_\xi\phi
= \pounds_\xi (F\d\phi) -\p_\chi F\pounds_\xi\chi \d\phi + F\pounds_{\d\xi}\phi,
\ee
where we used $[\d,\d_\xi]=0$ in the first equality, and $[\d,\pounds_\xi]=\pounds_{\d\xi}$ in the last, and the definition $I_{\d\xi}\d\phi:=\pounds_{\d\xi}\phi$.
From this formula we see that anomaly-freeness means
\be\label{Fanomalyfree}
\p_\chi F\pounds_\xi \chi=0.
\ee
Namely, $F$ should either not depend on the background fields $\chi$, or if it does, the symmetry group should be made only of symmetries of $\chi$. 
The symmetry group is typically required to preserve some universal background structure. If this is described by the background fields $\chi$, then the symmetry group  coincides with those diffeomorphisms that are symmetries of $\chi$, and there are no anomalies. 
But if the background structure is described equivalence classes of background fields, the situation is different, because isometries of the background structure need not be symmetries of individual representative fields, which are the quantities entering \eqref{Fanomalyfree}. 
We also see that the notion of covariance given by matching Lie derivatives can be stated equivalently as
\be\label{pFcovariant}
\p_\chi F\pounds_\xi \chi=F\pounds_{\d\xi}\phi. 
\ee
It means that representatives of the equivalence class for which the symmetry group is not an isometry can still be allowed, provided it carries a specific field-dependence. Even though it seems natural to talk about covariance when the two Lie derivatives match, it is the notion of anomaly-freeness that carries the most direct interpretation in terms of independence from background structures.
We will come back to this difference in Section~\ref{fielddep} at the end.

On a null boundary, $\Phi$ and $(g^{\Phi\Phi}:=g^{\m\n}\p_\m\Phi\p_\n\Phi)|_\cN$ are background fields.
Their anomalies are 
\be\label{anos}
\D_\xi \Phi = -\pounds_\xi \Phi = -\xi^\Phi, \qquad 
\D_\xi g^{\Phi\Phi} = -\pounds_\xi  g^{\Phi\Phi} 
\eqonN -\xi^\Phi\p_\Phi  g^{\Phi\Phi},
\ee
and vanish if we restrict attention to diffeomorphisms that satisfy $\xi^\Phi\eqonN 0$.
These are the tangent diffeomorphisms, and don't move the boundary. 
They can be parametrized as 
\be
\xi = \xi_{\sscr{\cN}}^a\p_a +\Phi \bar\xi^\m\p_\m \in{\Diff}(\cN), \qquad \xi\cdot l \eqonN 0,
\ee
with $\bar\xi$ smooth on $\cN$. We will refer to $\bar\xi^\m$ as the \emph{extension} of the symmetry vector outside of the boundary and into the bulk, and to the specific component $\xiext\eqonN-(f\Phi)^{-1}\xi\cdot l$ as the transversal extension. 
Since $\pounds_\xi g^{\Phi\Phi}\eqonN 0$ for $\xi\in T\cN$, we can write covariantly
\be
\hat \xi := \int \pounds_\xi g_{\m\n}\f{\d}{\d g_{\m\n}},
\ee
without the need to treat separately $g^{\Phi\Phi}$ and the dynamical components of the metric.

However, anomalies are present even for tangent diffeomorphisms if we have to deal with normal derivatives of the background fields. This is precisely the case at hand, since the pull-back of the symplectic potential depends on the normal 1-form. For a tangent diffeomorphism, we have
\be\label{defw}
\D_\xi l_\m \eqonN - w_\xi l_\m, \qquad \D_\xi l^\m \eqonN - w_\xi l^\m, \qquad w_\xi := (\pounds_\xi - \d_\xi )\ln f+\xiext.
\ee
We see that the anomaly $w_\xi$ depends on both the choice of normal representative, through the non-covariance of $f$, and on the diffeomorphism considered, through the transversal extension of the symmetry vector field, namely its $\Phi$ component.
As far as both quantities are arbitrary, one can choose them so that anomalies are vanishing, 
for instance taking $f=1$ and $\xiext=0$. However, while the choice $f=1$ is always acceptable (but may not be the best choice to study a specific problem), $\xiext=0$ is not, 
because in most cases of interest this extension is fixed to a non-vanishing value determined by the parameters $\xi^a_{\sscr\cN}$. These include the case of isometries, asymptotic symmetries at future null infinity, and 
 it would exclude from the symmetries the possibility of a Killing vector, whose transversal extension is fixed and non-vanishing. 
More in general, asymptotic symmetries at future null infinity as well as on a physical null hypersurface and a non-expanding horizon all require to fix the transversal extension of the symmetry vectors non a non-vanishing value determined by the parameters $\xi^a$. We will review below why.

To give further intuition about the meaning of $w_\xi$, consider the case of a non-null boundary. We still have \eqref{anos}, hence anomalies only appear for quantities like the normal, once we restrict attention to diffeomorphisms that are tangent to the boundary. For an arbitrary normal, \eqref{defw} is also still valid. But if we choose a unit-norm normal, then $w_\xi$ vanishes identically. If we recall that a unit-norm normal has the property of being independent of the embedding of the boundary, we see that anomalies arise not so much from the presence of a boundary, but rather from a foliation-dependence in its description. 
In other words, the equivalent of class-III invariance in the time-like case is achieved through invariance under $\Phi$ reparametrization only, because $f$ is fixed.
Coming back to the case of a null boundary, there is no foliation-independent description, and no canonical normalization for the normal, hence anomalies become relevant.

For the sake of this paper, \eqref{defw} is the main anomaly that we have to worry about, but not the only one.
A second source of anomalies is the rigging vector, which is also a non-dynamical and background quantity. Its anomalies are less important in the end, but will appear in some intermediate calculations and it is useful to track them as well. For an arbitrary choice of rigging, 
\be\label{anorig}
\D_\xi n_\m=w_\xi n_\m+Z_\m, \qquad Z \cdot l = Z \cdot n=0,
\ee
where the proportionality to $w_\xi$ of the first term follows from $l\cdot n=-1$, and the vector $Z$ parametrizes the rigging anomaly. Its explicit form depends on the specific choice of $n$, and we can leave it unspecified in the following.
For instance, the projector $\g_{\m\n}$ is manifestly class-III invariant but not class-I invariant. It has an anomaly determined by \eqref{anorig} as
\be\label{anogamma}
\D_\xi \g_{\m\n} = 2l_{(\m}Z_{\n)}.  
\ee

The anomalies \eqref{defw} and \eqref{anorig} correspond to the non-invariance under infinitesimal class-I and class-III transformations with parameters $A=e^{-w_\xi}\simeq 1-w_\xi$ and $a=m\cdot Z$.
Further anomalies appear for quantities with a non-vanishing spin weight, since these depend on the background structure $m$ associated with the choice of $n$. However these will not be relevant for us, we will always compute anomalies of quantities that can be expressed in terms of $l$ and $n$ alone.
For these, it is easy to prove that class-I and class-III invariance implies anomaly-free, see Appendix~\ref{AppAnomaly}. 

A subtle point to highlight is that anomaly-freeness requires class-III invariance in the general sense of a field-dependent rescaling. 
For instance, $n^\m \d l_\m$ is manifestly class-III invariant if the rescaling is field-independent, but not otherwise: $n^\m \d l_\m\to n^\m \d l_\m-\d \ln A$.
It is in fact anomalous, 
\be\label{Dndl}
\D_\xi (n^\m\d l_\m) = -\D_\xi \d \ln f = \d w_\xi -w_{\d\xi}.
\ee
We remark for later use that this 
specific anomaly vanishes if the variations are restricted by $\d l_\m\eqonN 0$, since $\xi$ is tangent to $\cN$. But it vanishes also if $\d l_\m\neq 0$ provided that $\d l^\m\eqonN 0$. This may not look obvious, but it follows from \eqref{defw}, and can be deduced also looking at \eqref{detidentity}.

Similarly, a quantity that is only partially class-III invariant like $\bar k\eps_\cN$
(we recall that it is independent of $f$ but not of invariant under reparametrizations of  $\Phi$)
 is also anomalous, 
\be\label{barkan}
\D_\xi (\bar k \eps_\cN) = -\pounds_l \xiext \, \eps_\cN.
\ee

\subsection{Anomalies of the boundary Lagrangians}

As a first application of this formalism, we compute the anomaly of the boundary Langrangians \eqref{ellfamily}. 
Using \eqref{defw} we find
\begin{align}
\D_\xi\eps_\cN=w_\xi\eps_\cN, \qquad \D_\xi\th=-w_\xi\th, \qquad \D_\xi\eps_S=0,
\qquad \label{anok}
\D_\xi k= -(\pounds_l+k)w_\xi. 
\end{align}
From the last one, we also deduce that
\be\label{liew}
\pounds_l w_\xi =(\pounds_\xi-w_\xi)k -\d_\xi k.
\ee
Adding up these contributions we have
\be\label{anellfamily}
a^{\sscr (b,c)}_\xi := \D_\xi \ell^{\sscr (b,c)} = b\pounds_l w_\xi \, \eps_\cN =b\, d w_\xi \w \eps_S, 
\ee
where in the last equality we used \eqref{defepsS} and the fact that $w_\xi$ is only defined on $\cN$.
As expected, any member with $b\neq 0$ is not class-III invariant and it is anomalous. 
The family of covariant boundary Lagrangians is \eqref{ellfamily} with $b=0$ and $c$ arbitrary. This includes in particular the Conf and CFP choices. In the latter case, notice that the statement about covariance is valid also if $\d k\neq 0$. 
The anomalous Lagrangians include the Dirichlet choice with $b=2$.
Such anomalies would only vanish in the special case $\pounds_l w_\xi=0$. Looking at \eqref{liew}, we see that this would occur for instance if the phase space is restricted to satisfy $k=\d k=0$.
Finally for the excluded member \eqref{ThbarD}, we have
\be\label{ellbaran}
a^{\sscr\bar D}_\xi := \D_\xi \ell^{\sscr \bar D} = 2\pounds_l \xiext \, \eps_\cN,
\ee
which captures explicitly its dependence on reparametrizations of $\Phi$.

This result shows that Dirichlet boundary conditions require an anomalous boundary Lagrangian, whereas conformal boundary conditions admit a covariant one.
The situation is the same if we move the expansion term to the corner and work with the primed boundary Lagrangians. In this case the conformal boundary Lagrangian vanishes and its covariance is obvious. Recalling the earlier discussion on ambiguities, we see that the anomaly here keeps track of the dependence of the Dirichlet boundary Lagrangian on non-geometric structures, hence of its ambiguity.
This comes from its dependence on the inaffinity and failure of being class-III invariant, and the problem is resolved switching to the conformal polarization instead.

Lagrangian anomalies appear in the study of central extensions of charge algebras.
As shown in  \cite{Chandrasekaran:2020wwn,Freidel:2021cjp}, the Lagrangian anomaly can be used to compute the cocyle that appears on the right-hand side of the Barnich-Troessaert bracket \cite{Barnich:2011mi}. This approach was applied in \cite{Chandrasekaran:2020wwn} to investigate whether one can obtain a central charge  that would be relevant to understand black hole entropy as proposed in \cite{Haco:2018ske}.
It was found that one can indeed reproduce the functional dependence of the entropy on the horizon's area, but with a wrong numerical factor. The right numerical factor would require as boundary Lagrangian $k\eps_\cN$ instead of $2k\eps_\cN$.\footnote{
This alternative boundary Lagrangian corresponds to the boundary condition $\d k=(\th+k)\d \ln\varepsilon$. They do not impose $\d l^\m=0$ however, instead the vector fields considered in \cite{Haco:2018ske,Chandrasekaran:2020wwn} stem from asymptotic symmetries of an auxiliary AdS$_3$ space that appears under a special coordinate transformation of the near-horizon geometry.}

\subsection{Anomalies of the symplectic potentials}\label{SecAnTh}

Next, we look at the anomalies of the symplectic potential. The standard symplectic potential $\th^{\sscr EH}$ is manifestly anomaly-free, since it depends only on the metric and its derivatives. So does its pull-back, since the anomaly operator commutes with taking the pull-back for tangent diffeomorphisms.
Decomposing it as in \eqref{ThN1} introduces the background structure given by the reference NP tetrad used, which captures the choice of normal representative and of rigging vector. This step does not introduce anomalies, since as we proved \eqref{ThN1} is both class I and III invariant.

Anomalies can instead appear in the preferred choice of $\th'$. From \eqref{th'} we have
\be
0=\D_\xi \th' - \D_\xi \d \ell +d\D_\xi \vth.
\ee 
Using this formula we can easily compute the anomaly for the family \eqref{ellfamily}. 
The corner term is \eqref{vthEH} in all cases, and its anomaly is given by
\be\label{anovth}
\D_\xi\vth^{\sscr EH} = 2\D_\xi (n^\m\d l_\m) \eps_S. 
\ee
Using this and \eqref{anellfamily}, 
\begin{align}\label{anoThbc}
& \D_\xi \th^{(b,c)} = b\, d w_\xi\w\d \eps_S +(b-2)d\D_\xi (n^\m\d l_\m)\w\eps_S- 2\th \D_\xi (n^\m\d l_\m) \eps_\cN.
\end{align}
The relation of this formula to the lack of class-III invariance is straightforwardly obtained with the replacement $\ln A=-w_\xi$.

For conservative boundary conditions with $\d\eps_S=0$, the anomaly vanishes for $b=2$ (the case of Dirichlet boundary conditions) if $\th=0$,
and for any $b$ and any $\th$ if $\D_\xi (n^\m\d l_\m)$ vanishes, which we recall follows from either $\d l_\m\eqonN 0$ or $\d l^\m\eqonN 0$.

What if we want to impose  leaky boundary conditions instead, with $\d \eps_S\neq 0$? We need either $b$ or $d w_\xi$ to vanish. 
The family with $b=0$ is covariant under a minimal set of restrictions:
\be
\d l^\m\eqonN 0 \quad {or}\quad \d l_\m\eqonN 0 \qquad \Rightarrow \qquad \D_\xi \th^{\sscr (0,c)}=0.
\ee
These are the most generic conditions that guarantee that the symplectic potential is anomaly-free for all $\xi$'s,
namely independent of the choice of normal representative.
If $b\neq 0$, we can use
\be\label{dwepsS}
d w_\xi\w\d \eps_S = \pounds_{\d l}w_\xi \eps_\cN+\pounds_l w_\xi \d \eps_\cN.
\ee
This vanishes if
\be\label{bcanofree}
\d l^\m=\d k=k\eqonN 0 \qquad \Rightarrow \qquad \D_\xi \th^{\sscr (b,c)}=0,
\ee
however with these conditions the terms in $b$ drop out completely. We conclude that the only relevant case is $b=0$.

We see that the conformal polarization is the only choice of \emph{diagonal} symplectic potential that is anomaly-free upon imposing only the minimal condition that $\d f=0$.
All  other choices considered are either not in diagonal form, or require additional restrictions on the phase space. This makes the  conformal polarization best suitable to study more general leaky boundary conditions without introducing anomalies. We will see below that anomalies in the symplectic potential spoil the integrability of Hamiltonian charges.

The anomaly of the symplectic potentials can also be derived summing up the anomalies of each spin pair, which we report here for completeness.
The shear tensor $\s^{\m\n}$ is neither class-I nor class-III invariant, and it carries both anomalies:
\be
\D_\xi \s^{\m\n}  = -w_\xi\s^{\m\n} + 2l^{(\m}\s^{\n)\r}Z_\r
\ee
(whereas the NP scalar being rigging-independent only carries the first anomaly, $\D_\xi \s=-w_\xi \s$).
Using this and  \eqref{anogamma} in the spin-2 pair of the symplectic potential, we find
\begin{align}\nn
 \D_\xi (\s^{\m\n}\d \g_{\m\n}\eps_\cN) &= \D_\xi(\s^{\m\n}\eps_\cN)\d\g_{\m\n} + \s^{\m\n}\eps_\cN \D_\xi\d \g_{\m\n}
 = 2l^{\m}\s^{\n\r}Z_\r\d\g_{\m\n} \eps_\cN \\\label{spin2Y}
&= \d l^\m(\th Z_\m -  2 Z^\r\na_\n l_\r) \eps_\cN ,
\end{align}
where we used $\D_{\d\xi}g_{\m\n}=0$ and the orthogonality properties of $Z$.
Next, we have
\begin{align}
 \D_\xi\pi_\m  = 2 (\p_\m + l_\m\p_n)w_\xi + 2Z^\r \na_\m  l_\r - \th Z_\m + 2l_\m(Z^\r n^\n +n^\r Z^\n)\na_\n l_\r, 
\end{align}
so the spin-1 pair has anomaly 
\begin{align}
\D_\xi(\pi_\m\d l^\m\eps_\cN) &= \D_\xi\pi_\m\d l^\m\eps_\cN -\pi\cdot l \,  (\d w_\xi-w_{\d\xi}) \eps_\cN
\nn\\&=(2\p_\m w_\xi + 2Z^\s \na_\m l_\s -\th Z_\m)\d l^\m\eps_\cN +(2k-\th)\D_\xi (n^\m\d l_\m) \eps_\cN.
\end{align}
where we used  $\pi\cdot l = \th-2k$. The last term in the RHS depends on $\d l_\m$ but vanishes for $\d l^\m\eqonN 0$ as discussed below \eqref{Dndl}.
Finally for the spin-0 part let us consider the cases of Dirichlet and conformal polarizations as examples.
\begin{align}
& - \D_\xi[(\th+2k)\d\eps_\cN] = - (\th+2k)\D_\xi (n^\m\d l_\m) \eps_\cN + 2\pounds_l w_\xi \d\eps_\cN.
\end{align}
Adding up the three contributions we recover \eqref{anoThbc} for $b=2$,
\be\label{anoThD}
\D_\xi \th^{\sscr D}=2d w_\xi\w\d \eps_S - 2\th \D_\xi (n^\m\d l_\m) \eps_\cN.
\ee
Switching to conformal polarization,
\begin{align}
\D_\xi[\d(\th+2k)\eps_\cN] = -(\th+2k)\D_\xi (n^\m\d l_\m) \eps_\cN - 2(\pounds_{\d l} w_\xi +\pounds_l \D_\xi (n^\m\d l_\m))\eps_\cN.
\end{align}
In the last term we can use
\be
\pounds_l \D_\xi (n^\m\d l_\m)\eps_\cN = d \D_\xi (n^\m\d l_\m)\w\eps_S.
\ee
Adding up we recover \eqref{anoThbc} for $b=0$,
\be\label{anoThY}
\D_\xi \th^{\sscr Conf} = - 2d[\D_\xi (n^\m\d l_\m) \eps_S].
\ee 
This derivation allows one to appreciate that the potential anomaly coming from the term $\d k$ present in $\th^{\sscr Conf}$ cancels out with a contribution coming from the spin-1 term. 
The subtle point is that even though the anomaly of $k$ does not vanish for $\d l^\m=0$, the anomaly of $\d k\, \eps_\cN$ does.
As a consequence, it is crucial in order for the anomaly to be a boundary term that the spin-1 momentum is $\pi_\m$, and not $\eta_\m$ only, as in \cite{Hopfmuller:2018fni}.
In fact the restriction $n_\m\d l^\m\eqonN 0$ they use is not class-I invariant, and introduces an anomaly in their symplectic potential that cannot be eliminated requiring $\d l_\m\eqonN 0$. In conclusion, we state again that the conditions for anomaly-freeness are $\d l_\m\eqonN 0$, and not other non-covariant restriction, or the whole $\d l^\m\eqonN 0$.

\subsection{Boundary symmetry groups}\label{SecSym}

In this Section we review the different boundary symmetry groups that have been considered in the literature, with emphasis on the different background structures kept fixed. We show how they can be derived in a simple way using $\d_\xi$ and $w_\xi$. We highlight how each additional restriction on the variations affects the symmetry group, the extension  of the symmetry vector fields, and the anomalies.

The minimal requirement that we consider is that the boundary should be a null surface, which restricts the variations to satisfy $l_\m\d l^\m\eqonN 0$. The residual diffeomorphisms that preserve the boundary and the condition that it is null must satisfy $\pounds_\xi \Phi\eqonN 0$ and 
\be\label{xinull}
l_\m\d_\xi l^\m = l_\m(\pounds_\xi l^\m+\D_\xi l^\m)\eqonN l_\m l_\n \pounds_\xi g^{\m\n}\eqonN 0.
\ee
These equations are solved by $\xi^\Phi\eqonN0$,
namely any diffeomorphism tangent to the boundary.  
For later convenience, we parametrize the tangent vectors in affine coordinates as
\be\label{xigen}
\xi = \t(\l,x^A)\p_\l +Y^A(\l,x^B)\p_A +\ldots
\ee
The restriction of these vectors to $\cN$ is arbitrary, hence they span the whole group $\Diff(\cN)$.\footnote{Here $\t$ is a free function and not the NP coefficient. NP notation will not be used in the rest of the paper.}
The dots here denote the extension of the vector field off the hypersurface, which is also arbitrary. 
This means that the anomaly \eqref{defw} is also arbitrary.

As we will see below, the gravitational charges depend on the symmetry vectors and their derivatives, and therefore on the extension. More precisely, the charges turn out to depend on the anomaly \eqref{defw}, which is determined by the choice of $f$ and the extension component $\xiext$. 
The relevant symmetry vector fields for the charges are thus $\xi^a\p_a+\Phi\xiext\p_\Phi$. They close under the Lie bracket and span the subgroup $\Diff(\cN)\ltimes \R^\cN$ parametrized by four free functions on $\cN$. Therefore one can conclude that $\Diff(\cN)\ltimes \R^\cN$ is the symmetry group of the most general phase space with arbitrary metric variations on a given null boundary. This group appears also in \cite{Adami:2021nnf}.
At this stage, the physical relevance of a symmetry group not intrinsic to the boundary is unclear to us. It means in particular that the charges can have arbitrary values even though the restriction of the vector field to the boundary hypersurface vanishes.\footnote{The case of isometries at the bifurcation surface is not a counter-example: it is true that there one gets a non-zero charge even though the restriction of the symmetry vector field to the bifurcation surface vanishes, but the restriction of the symmetry vector field to the 3d boundary, namely the horizon, does not.} 
For this reason we think that this enlarged phase space does not provide a good handle for the study of dynamical geometric properties of a null hypersurface.\footnote{A different viewpoint is taken in \cite{Chandrasekaran:2023vzb}, where it is argued that this additional free parameter should be taken seriously as a characterization of the near-null hypersurface geometry.}

Next, we add the restriction that the tangent vector be fixed, $\d l^\m\eqonN 0$, as done for instance in \cite{Lehner:2016vdi}. 
The residual diffeomorphisms must satisfy
\be\label{xilup}
\d_\xi l^\m\eqonN 0.
\ee
The analysis of these three conditions can be split in two cases, corresponding to null and space-like components:
\begin{align}\label{ndxil}
& n_\m \d_\xi l^\m = n_\m l_\n \pounds_\xi g^{\m\n} -\d_\xi \ln f \eqonN 0 && \Rightarrow \quad {\rm restricts \ the \ extension} \\
& m_\m\d_\xi l^\m = m_\m\pounds_\xi l^\m \eqonN 0 &&  \Rightarrow \quad {\rm restricts \ the \ allowed \ diffeos}
\end{align}
To understand the first condition, observe that in adapted coordinates it contains the term 
$\p_\Phi \xi^\Phi\eqonN\xiext$. This equation fixes the extension component $\xiext$, and the symmetry group is thus $\Diff(\cN)$. Having fixed part of the extension however makes closure under the spacetime Lie bracket non-trivial, and an additional condition is needed. We will come back to this shortly.
To understand the second restriction, it is easiest to take a coordinate system $(\l,\Phi, x^A)$ with $\l$ affine parameter, so that 
\eqref{minimalgf} holds.
Then 
\be\label{dxilaffine}
\d_\xi l^\m =(\d_\xi \ln f -\xiext)l^\m -f\p_\l\xi^\m 
\eqonN 0 \quad \Rightarrow \quad \left\{\begin{array}{lll} 
\m=\Phi &\quad \xi\in T\cN \\ \m=\l &\quad \xiext = \d_\xi \ln f - \p_\l\t \\ \m=A &\quad \p_\l Y^A=0
\end{array}\right.
\ee
The $\l$ component shows that in these coordinates the extension is restricted via a function of the time derivative of $\t:=\xi^\l$ and of $\d_\xi f$.\footnote{It should be clear that the equation can not be solved taking $f$ as a function of an arbitrary extension, because that would make the scaling of the normal dependent on the diffeomorphism considered.} 
Comparing with \eqref{defw}, we see that 
the anomalies for these residual diffeos read
\be\label{BMWanomaly}
w_\xi =\pounds_\xi \ln f- \dot\t.
\ee 
They are thus determined by the parameters of the symmetry vector field plus the choice of $f$.
The cross-section components $A$  show that the tangential diffeomorphisms must be time-independent, $\dot Y^A=0$.
Namely they become the $\Diff(S)$ super-Lorentz transformations encountered at future null infinity in \cite{Campiglia:2014yka,Compere:2018ylh}. In the following, we will refer to time-independent arbitrary diffeomorphisms of the cross-sections as super-Lorentz.
On shell of \eqref{dxilaffine}, the symmetry vector fields  \eqref{xigen} reduce to
\be\label{xiMyers}
\xi = \t(\l,x^A)\p_\l +Y^A(x^B)\p_A - \Phi(\dot\t-\d_\xi\ln f)\p_\Phi+\ldots,
\ee
where the dots include the part of the extension left arbitrary. 

These vector fields however do \emph{not} close under the spacetime Lie bracket, see Appendix~\ref{Lieclosure} for a proof. It happens only if we require $\d_\xi f=0$, or if $\D_\xi f=0$. The first option is a priori more general since it can be achieved without any further restriction on $\xi$ simply requiring $\d l_\m\eqonN 0$.
Having done so, the vector fields are given by
\be\label{xibelli}
\xi = \t \p_\l +Y^A\p_A - \Phi \dot\t \p_\Phi+\ldots, \qquad \t(\l,x^A), \quad Y^A=Y^A(x^B).
\ee
The extension $\xiext$ no longer depends on $f$, and is entirely determined by the parameters of the symmetry vector fields at $\cN$.
These vector fields close under the Lie bracket, and span the 
subgroup 
\be\label{bigG}
\bigG:= \Diff(S)\ltimes \Diff(\R)^S \subset \Diff(\cN).
\ee
This is the `little group' of diffeomorphisms preserving the null geodesic congruence spanned by $l$ on $\cN$.
The semi-direct product structure follows from \eqref{comm1}, and it means in particular that the identification of the group components with $Y^A$ and $\t$ is not canonical, but relies on the choice of affine coordinates we made. This is a situation familiar from the BMS and CFP groups.

There is also another way to understand the importance of adding the condition $\d l_\m\eqonN 0$. Without it, the four conditions \eqref{xinull} and \eqref{dxilaffine} imply only three  restrictions of the metric variations, because varying $f$ does not affect the metric. Therefore the residual diffeomorphisms \eqref{xiMyers} do not correspond to a complete gauge fixing like \eqref{nonaffinegf}.
Imposing $\d l_\m\eqonN 0$ turns the four conditions into four conditions on the metric variations, given by:
\be\label{ldg}
\d l^\m=\d l_\m\eqonN 0 \quad \Rightarrow \quad l^\m \d g_{\m\n} \eqonN 0.
\ee
Notice that the last equation is class-III invariant, hence the symmetry group satisfying $l^\m \d_\xi g_{\m\n} \eqonN 0$ depends only on the equivalence class of normals and not on a choice of representative. Diffeomorphisms preserving this condition satisfy $l^\m \pounds_\xi g_{\m\n}=0$. In affine coordinates, and restricting $\xi$ to be tangential, we get
\be
\xi^\Phi \eqonN 0, \qquad \p_\Phi\xi^\Phi+\p_\l\xi^\l\eqonN 0, \qquad \p_\l\xi^A\eqonN0,
\ee
that coincide with the restriction of \eqref{dxilaffine} to $\d_\xi f=0$.\footnote{Notice that it is necessary to restrict upfront to tangential diffeomorphisms, otherwise one obtains the larger set of solutions with $\xi^\Phi\eqonN f(x^A)$. We also point out that the weaker set $l^\m \d g_{\m\pbi{\n}}  = 0$ misses $\p_\l \xi^\m+\p_\Phi\xi^\Phi=0$, namely \eqref{ndxil}.}

Let us now see the effect of adding $\d k=0$ on top of the previous conditions. 
From \eqref{dkspecial} we have
\be\label{dxikspecial}
\d_\xi k = - \f12 n^\m\na_\m (l_\n l_\r \pounds_\xi g^{\n\r})\eqonN 0.
\ee
This equation involves the first derivatives off the hypersurface of the metric. What it does is to further restrict the diffeomorphisms to preserve the condition of affine coordinates, namely the metric in the form \eqref{minimalgf}, as opposed to the more general form \eqref{nonaffinegf}.
It is easy to see that in affine coordinates the equation simplifies to 
\be\label{ddottau}
\ddot \t=0. 
\ee
This means that we can write $\t=T(x^A)+\l W(x^A)$ in terms of a supertranslation with parameter $T$ and a Weyl transformation with parameter $W$. 
We have thus recovered the symmetry group of \cite{Chandrasekaran:2018aop}, 
\be\label{GCFP}
G^{\sscr CFP}:=(\Diff(S)\ltimes \R^S_W)\ltimes \R_T^S.
\ee
We will refer to this group as CFP from the authors of  \cite{Chandrasekaran:2018aop}. An alternative good name is BMSW group. Its vector fields read
\be\label{xiCFP}
\xi = T\p_\l +Y^A\p_A +W(\l\p_\l- \Phi\p_\Phi)+\ldots. 
\ee
in affine coordinates
As for the anomaly, this is still given by \eqref{BMWanomaly}, now
\be\label{anaffine}
w_\xi = T\p_\l\ln f+W(\l\p_\l\ln f-1)+Y^A\p_A \ln f.
\ee
To be precise, this is the symmetry group if $\cN$ is complete. If it is semi-complete instead, super-translations must be dropped because they do to preserve the boundary of $\cN$ \cite{Chandrasekaran:2018aop}. This happens for instance if $\cN$ is the boundary of a causal diamond \cite{Chandrasekaran:2019ewn} or a light-cone. If $\cN$ has two boundaries, as for example when connecting two space-like surfaces, then $\t$ must vanish at two different values of $\l$. This removes both super-translations and Weyl rescaling from the CFP group, which reduces in this case to $\Diff(S)$ only. The situation is different for the ${\Diff_l(\cN)}$ group, which having arbitrary time-dependence in the super-translations survives as a symmetry group with $\Diff(S)$ plus a left-over super-translations with $\t(\l_0)=\t(\l_1)=0$ (which do form a group) also for a $\cN$ with two boundaries. This is a physical set-up in which the new group ${\Diff_l(\cN)}$ is important.

If the covariant phase space is restricted to describe only NEHs, then the background structure can be  strengthen even more to allow for a constant rescaling only of the normal, and then the group is restricted to constant $W$'s and conformal isometries of the sphere \cite{Ashtekar:2021kqj},
\be
G^{\sscr AKKL}:=(\SL\ltimes \R_T^S)\times \R^+_W.
\ee
We will refer to this group as AKKL from the authors of  \cite{Ashtekar:2021kqj}. An alternative good name is NEH group.

Let us comment on the 
method we used to derive the symmetry groups $\bigG$ and $G^{\sscr CFP}$. This was based on identifying the diffeomorphisms that preserve the variations required to vanish, as opposed to the more common approach in the literature that consists on identifying the isometries of the background structure. 
But it is easy to prove the equivalence of our method in this context. 
From \eqref{defw}, we see that $\d_\xi l^\m\eqonN 0$ is equivalent to $\pounds_\xi l^\m \eqonN w_\xi l^\m$, namely the diffeomorphisms that preserve the equivalence class $[l^\m=Al^\m]$. Same story for $\d_\xi l_\m\eqonN 0$. For  $\d_\xi k\eqonN 0$, we see from \eqref{anok} that is equivalent to $\pounds_\xi k \eqonN w_\xi(k+\pounds_l\ln w_\xi) $, namely the diffeomorphisms that preserve the equivalence class \eqref{CFPclass}.
The three equations for the Lie derivatives of $l^\m, l_\m$ and $k$ are indeed the conditions used in \cite{Chandrasekaran:2018aop}. 

We also remark the
importance of using affine coordinates in order to solve for the vector fields that satisfy the phase space restrictions. The conditions \eqref{xilup} and \eqref{dk0} are in fact  complicated in arbitrary coordinates, and boil down to simple statements about time-independence in affine coordinates.
More importantly, the $\xi$'s solving these equations in arbitrary coordinates depend explicitly on the metric,
and can be characterized in a metric-independent way only in affine coordinates.
What makes affine coordinates special is that they corresponding to a gauge fixing whose preservation coincides with preserving the background structure. Otherwise one cannot describe the symmetry vector fields in a universal way, and must work with field-dependent diffeomorphisms.\footnote{This observation
should be contrasted with the analysis in the CFP paper, where the vector fields where characterized in a metric-independent way independently of the choice of coordinates. We believe that the reason for this is that their characterization is done directly in terms of intrinsic quantities on the hypersurface only, and therefore in a metric-independent way (as shown by \eqref{ldg}, it is only by looking at spacetime restrictions that the symmetry group can be seen as preserving a metric gauge-fixing).  We also remark that when they construct a spacetime diffeomorphism representative of the symmetry, they \emph{define} it to match the intrinsic diffeomorphism when restricted to $\cN$, see their (5.6). We can do the same here: once the metric-dependent vector fields are found in arbitrary coordinates, we can do an intrinsic diffeomorphism on $\cN$ that maps them to the metric-independent one in affine coordinates.}

How about the Robin-type condition \eqref{Ybc}? In this case, \eqref{dxikspecial} is replaced by
\be\label{Robinxi}
\d_\xi k = - \f12 \pounds_l(\na\cdot\xi), 
\ee
which in affine coordinates becomes
\be\label{tauRobin}
-2\ddot \t = \p_\l(\t \th_\l+D_AY^A) = \dot\t \th_\l+\t\dot\th_\l +Y[\th_\l],
\ee
where $D_A$ denotes the derivative on the cross-section, and $\th_\l$ is the expansion of the affine normal.
The RHS vanishes for a NEH, but not in general. 
The novelty with respect to the previous cases is the metric-dependence of the equation in affine coordinates.
We don't know the general solution to this equation, but it is clear that it will be a metric-dependent function 
$\t=\t(\l,T,W,Y;g)$, where $T(x^A)$ and $W(x^A)$ are integration constants. 
These vector fields are labelled by the same parameters  space as \eqref{GCFP}, but they are field-dependent. 
This case will be studied elsewhere.

We have shown what happens when one implements one by one the vanishing of $n_\m \d l^\m$, $\d l^\m$, $\d l_\m$ and $\d k$, and how it determines the symmetry groups $\Diff(\cN)\ltimes \R^\cN$, $\Diff(\cN)$, $\bigG$, $G^{\sscr CFP}$, and their anomalies. We proceeded in this specific order, because it is the one that appears the most useful to us, but with the same method one can consider any mixture of partial implementations. Let us briefly comment on a few of these partial alternatives.

The condition \eqref{ddottau} can be derived also without imposing the condition $\d l_\m\eqonN 0$. To see that, 
we start from the general formula \eqref{dkgen}. Restricting to $\d l^\m\eqonN0$, we have
\be
\d k= \f12 l^\m l^\n n^\r \na_\r \d g_{\m\n} -n^\m l^\n l^\r \na_\r \d g_{\m\n}.
\ee 
The residual diffeomorphisms preserving this condition must satisfy
\be\label{dk0}
\d_\xi k = l^\m l^\n n^\r (R^\s{}_{\m\n\r}\xi_\s-\na_\m\na_\n\xi_\r)\eqonN 0.
\ee
The term in bracket can be recognized as a property of a Killing vector, but the allowed $\xi$'s are here more general since only a specific scalar contraction of that term is being imposed to vanish, and on the hypersurface only. The $\xi$'s solving this equation are complicated functions of the metric in general, but 
in affine coordinates it gives back \eqref{ddottau}.
As explained earlier though, in the extended phase space with $\d l_\m\neq 0$ the vector fields don't close under the spacetime Lie bracket, and don't correspond to the residual gauge fixings preserving \eqref{minimalgf}.

Suppose now that we require $\d l_\m\eqonN 0$ and/or $\d k=0$ without fixing $l^\m$. Imposing $\d_\xi l_\m\eqonN0$ can only be solved if $\d f=0$, and this imposes no restriction on the symmetry vector fields. We have $\Diff(\cN)\ltimes \R^\cN$ with an arbitrary extension $\xiext$ and general anomaly  \eqref{defw}, and if we add $n_\m \d_\xi l^\m=0$ we have $\Diff(\cN)$ with anomaly \eqref{BMWanomaly}.

Preserving $k$ on top of $\d_\xi l_\m\eqonN0$ leads to
\be
\d_\xi k\eqonN 2 l_\r \na^{(\m} \xi^{\r)} n^\n\na_\m  l_\n  - l^\m l^\n n^\r \na_\r \na_\m\xi_\n.
\ee
This equation is not  class-III invariant, therefore the diffeos solving this equations depend on the choice of normal, and cannot be characterized in purely geometric terms. For instance for a null diffeo $\xi=\t \p_\l$, we get $k(\pounds_l+k)\t/f=0$, which is solved by any $\t$ for $k=0$, and by $\dot\t=(\dot f-k)\t/f$ for $k\neq 0$. The situation is similar if we preserve $k$ without preserving $l_\m$, just the above equation become more complicated, and remains not class-III invariant.
We conclude that imposing $\d k\eqonN 0$ without $\d l^\m\eqonN 0$ leads to symmetry groups which depend on structures unrelated to the geometry of $\cN$.

For related leaky boundary conditions and symmetry groups see also \cite{Donnay:2015abr,Donnay:2016ejv,Grumiller:2019fmp,Compere:2019bua,Adami:2020amw,Fiorucci:2021pha,Adami:2021nnf,Adami:2021kvx}.

\section{Charges and fluxes}

Having discussed the different symmetry groups associated with a larger or smaller background structure, we now briefly review how the covariant phase space allows one to associate Noether charges and 
Hamiltonian generators to these symmetries.
In particular, we will discuss how the choice of polarization affects the definition of charges and their fluxes, and what is the role played by anomalies. 

In the context of the covariant phase space, \eqref{th'} represents the freedom in choosing the symplectic potential: $\ell$ comes from the freedom of adding a boundary term to the Lagrangian without affecting the field equations, and $\vth$ from the fact that the Lagrangian only determines the symplectic potential up to an exact 3-form. We refer to the choice of $\th$ that can be read directly from $\d L$ without any additional information as the standard, or bare, choice.\footnote{This bare choice can also be mathematically selected if one defines the potential using the homotopy operator \cite{Iyer:1995kg,Barnich:2001jy,Freidel:2020xyx}.}

If one starts from a covariant Lagrangian $L$ and a covariant symplectic potential one obtains the following well-known formulas \cite{Iyer:1994ys} 
for the Noether current $j_\xi$,
\be \label{dq1}
j_\xi:= I_\xi\th-i_\xi L\eqons dq_\xi, \qquad d j_\xi \eqons 0,
\ee
and for the infinitesimal Hamiltonian generator $h_\xi$,
\be \label{Ixiom}
\sd h_\xi := -I_\xi\om = \d I_\xi\th - d i_\xi \th \eqons d(\d q_\xi -q_{\d\xi} - i_\xi\th).
\ee
Here we used the exterior calculus notation as in \cite{Freidel:2021cjp}, $\om:=\d\th$ is the symplectic 2-form current,  $\eqons$ means on-shell of the field equations and $q_\xi$ is the Noether surface charge.
A flux-balance law for this charge can be derived taking the pull-back of \eqref{dq1} along a lateral boundary (time-like or null). This eliminates the last term when $\xi$ is restricted to a symmetry vector, since the latter is tangent to the boundary. It follows that $I_\xi\th$ is the flux determining the variation of the charge along that boundary. 

There are however two problems with \eqref{dq1}. The first is that the associated Noether charges may be of 
little practical use. For instance with the bare choice $\th=\th^{\sscr EH}$ given by \eqref{ThEH}, the resulting Noether charges are given by the Komar 2-form \cite{Iyer:1994ys}.
These have notorious problems such as wrong numerical factors, and not being conserved even in the absence of radiation.
The second problem is that they don't coincide in general with the Hamiltonian generators, as one can see from \eqref{Ixiom}. This discrepancy is referred to as the problem of integrability of the infinitesimal generator, see e.g. 
\cite{Barnich:2001jy,Barnich:2010eb,Barnich:2011mi,Compere:2018ylh,Compere:2020lrt}.

The problems can be addressed using the Wald-Zoupas procedure \cite{Wald:1999wa}, 
which aims at prescribing a (possibly unique) set of charges requiring them to coincide with the canonical generators when a physically identified flux vanishes. This preferred flux is selected from the equivalence class \eqref{th'} based on covariance and physical criteria, for instance such that the charges 
are constant under conservative boundary conditions, or for perturbations around special solutions corresponding to stationary spacetimes. 
We then showed in \cite{Odak:2022ndm} how this procedure can be extended to include corner contribution, anomalies and field-dependent diffeomorphisms.
We also showed under which conditions the resulting WZ charges can be identified as Noether charges for a specific choice of boundary Lagrangian (see also \cite{Harlow:2019yfa,Chandrasekaran:2020wwn,Freidel:2021cjp,Chandrasekaran:2021vyu} on this).

Let us briefly recap some details of the WZ procedure as extended in \cite{Odak:2022ndm}.
We consider here only field-independent diffeomorphisms, because this is sufficient to understand the symmetry groups described in the previous Section. The case with field-dependent diffeomorphism is discussed at the end.
Starting from $\th=\th^{\sscr EH}$ and a given hypersurface, we select a preferred symplectic potential $\th'$ in the equivalence class \eqref{th'} satisfying three criteria:

\bit
\item[0.] $\d\vth=0$, so that $\om'=\om$;
\item[1.] $\D_\xi \th'=0$, so that the preferred potential is anomaly-free;
\item[2.] $\th'$ is in the form $p\d q$ where:
\smallskip \\
Case I: $\d q=0$ for conservative boundary conditions, which imply that $\om'$ has vanishing pull-back on the boundary;
\smallskip \\ Case II: $p=0$ for points in phase space satisfying a useful notion of stationarity, and the pull-back of $\om'$ is non-vanishing.
\eit
Ideally, these criteria should be enough to select a unique $\th'$.\footnote{This uniqueness may also require fixing field-space constant terms using a special solution as reference. In the following we neglect the discussion of these constant terms.} 
If this fails, additional or revisited conditions should be considered.
Notice that one typically  imposes some boundary conditions also in case II, weaker than the conservative ones, and needed to preserve  a certain boundary structure of physical relevance, for instance in order to characterize graviational radiation. The boundary structure shared by a certain class of metrics is referred to as their \emph{universal structure}. The conservative boundary conditions of case I 
are of the same type that are used in the variational principle. We refer to the generic, weaker set of boundary conditions of case II as leaky.

If the preferred $\th'$ and its Lagrangian $L'=L+d\ell$ are covariant, then the formulas \eqref{dq1} and \eqref{Ixiom} are still valid with primes everywhere. 
The importance of the condition 2 is then clear: when $\th'$ vanishes, the Noether charges coincide with the canonical generator for field-independent diffeomorphisms. Furthermore, they are automatically conserved in the subset of the phase space satisfying the conditions of case I or II.
The new Noether charges are related to those of $\th$ by \cite{Harlow:2019yfa,Freidel:2020xyx,Freidel:2021cjp}
\be\label{q'}
q'_\xi=q_\xi+i_\xi\ell-I_\xi\vth,
\ee
and are sometimes referred to as boundary-improved, or improved for short, Noether charges.

However, there is a caveat. In spite of the covariance requirement of condition 1, the selection process may introduce anomalies. This happens  
if the preferred $\th'$ is associated to a new Lagrangian $L'=L+d\ell$ whose boundary term is anomalous: $a'_\xi:=\D_\xi \ell\neq 0$.
In this case the new charges do not satisfy Noether's theorem in its original form \eqref{dq1}, because 
that relies on the covariance of $L$, specifically on the fact that $\d_\xi L=\pounds_\xi L=di_\xi L$. If the boundary Lagrangian is anomalous we have instead $\d_\xi L'=d(i_\xi L'+a_\xi)$ and the formula becomes
\be\label{dq2}
j'_\xi:= I_\xi \th' -i_\xi L' -a'_\xi\eqons d q'_\xi,\qquad d j'_\xi \eqons 0.
\ee
Condition 2 is no longer sufficient to guarantee the conservation of the Noether charges $q'_\xi$. 
Even worse, for a generic anomaly $a'_\xi$ we are not even guaranteed that the charges are conserved for isometries.
This potential problem is avoided thanks to condition 1. 
In fact, the pull-back of the Hamiltonian generators on the lateral boundary gives
\be\label{Ixiom2}
-\pbi{I_\xi\om} = \d I_\xi\th' -(\d_\xi -\pounds_\xi )\th' - d i_\xi \th' \eqons \d(dq'_\xi+a'_\xi) -(\d_\xi -\pounds_\xi )\th' -di_\xi\th'.
\ee
Here we used condition 0 and $d\th'\equiv 0$ (that follows since $\th'$ is only defined after pull-back) in the first equality, and \eqref{dq2} in the second.
If condition 1 holds
the generator is integrable once the preferred flux is subtracted:
\be\label{Ixiom3}
-I_\xi\om+di_\xi\th' = \d I_\xi\th'\eqons \d(dq'_\xi+a'_\xi).
\ee 
Furthermore, condition 1 also implies that  the Lagrangian anomaly must be spacetime-exact, specifically that $a'_\xi=ds_\xi$ where 
$\d s_\xi=-A'_\xi$ and $A'_\xi$ is the symplectic anomaly of the preferred $\th'$ \cite{Freidel:2021cjp}.
This makes it possible to define the WZ charges 
\be\label{WZcharges}
q^{\sscr WZ}_\xi:=q'_\xi+s_\xi.
\ee
They satisfy the flux-balance laws
\be\label{WZfluxes}
dq^{\sscr WZ}_\xi \eqons I_\xi \th', \qquad d \d q^{\sscr WZ}_\xi\eqons -\pbi{I_\xi\om}+di_\xi\th' .
\ee
It follows that they are conserved \emph{and} provide Hamiltonian generators when $\th'$ vanishes, be it for conservative boundary conditions, or leaky boundary conditions around stationary configurations. 

This is the Wald-Zoupas prescription. We stress that its keystone is condition 1. Without condition 1, 
we would in fact be stuck with \eqref{dq2}, without the possibility to use \eqref{Ixiom2} to justify and be guaranteed that the anomaly can be reabsorbed in the definition of the charge. With $a'_\xi$ (or even just part of it) still on the LHS of \eqref{dq2}, stationarity of $\th'$ would fail to give conserved charges, hence condition 2 would entirely lose its physical relevance. 

We also stress that even if the selected $\th'$ is covariant and the Lagrangian anomaly $a'_\xi$ drops out of the flux-balance laws  in the end, anomalous transformations can still be present, since
\be
I_\xi \th' = p\pounds_\xi q +p\D_\xi q.
\ee
This anomaly contribution is physically correct, because it is the right quantity to have so that background structures don't contribute to the flux: remember in fact that $\d_\xi l^\m\eqonN 0$ but $\pounds_\xi l^\m\neq 0$, for instance.

A natural question at this point is whether the WZ charges \eqref{WZcharges} can always be interpreted as Noether charges like \eqref{q'} for some boundary Lagrangian. The answer is yes iff $s_\xi = -\D_\xi c$ for some local 2-form $c$ constructed out of the fields and the background structure, and in this case the correct boundary Lagrangian is the anomaly-free choice $\ell+dc$  .
So we have two approaches to constructing charges: the Noether charge, based on selecting specific $\th'$ and $\ell$. And the WZ prescription, based on selecting only a preferred $\th'$. The convergence of the two approaches is obtained when the WZ charges can be derived as improved Noether charges with the choice of $\ell$ determined by a condition of covariance \cite{Odak:2022ndm}. 

An important point to appreciate is that the charges are anomalous even if the symplectic potential is not. This is simply because the charges depend on the symmetry vector fields $\xi$ which are generally anomalous. What one should require then is that the charge anomaly is sourced only by the $\xi$'s, namely that 
\be\label{goodcharge}
\D_\chi q_\xi = \f{\p q_\xi}{\p \xi}\D_\chi\xi = - \f{\p q_\xi}{\p \xi} \pounds_\chi\xi = - q_{[\chi,\xi]}.
\ee
This is precisely what is guaranteed for the WZ charges thanks to the covariance requirements of symplectic potential and boundary Lagrangian. 

It is also possible to drop condition 0, and consider a generalized WZ prescription based on 1 and 2 alone  \cite{Harlow:2019yfa,Odak:2022ndm}. 
This generalization will not be needed here, but it is necessary in order to obtain Brown-York charges at finite time-like boundaries with non-orthogonal corners \cite{Harlow:2019yfa,Odak:2021axr}, and for the generalized angular momentum of $\Diff(S)$ at future null infinity \cite{Compere:2018ylh}.
The WZ charges are still given by \eqref{WZcharges}, but where \eqref{q'} has a non-trivial $\vth$ term, and $\om'$ replaces $\om$ in \eqref{WZfluxes}. 

Summarizing, our viewpoint as put forward in \cite{Odak:2022ndm} is that the crux of the WZ procedure is really condition 1 (or its alternative version as anomaly-freeness). Condition 0 can be dropped, and condition 2 should be interpreted as a framework rather than a unique set-up, meaning that different notions of conservative boundary conditions or stationarity conditions can be considered, in order to describe different physical problems.

\subsection{Wald-Zoupas conditions on null hypersurfaces}

%
We now study which of the family of symplectic potentials $\th^{\sscr (b,c)}$ in \eqref{Thbc} 
satisfies the WZ conditions. We will recover the results of \cite{Chandrasekaran:2018aop},  show how they change for different polarizations, and how they can be extended to the relaxed phase space with $\d k\neq 0$.

Condition 0. Looking at \eqref{vthEH}, we see that it requires 
\be\label{cond0sat}
\d l^\m= \d l_\m \eqonN 0.
\ee
These can be satisfied without any restriction on the dynamics, and correspond to the gauge fixing \eqref{ldg}.

Condition 1. 
From Section~\ref{SecAnTh} we know that the family with $b=0$ and $c$ arbitrary is covariant for either one of the two conditions in \eqref{cond0sat}.

Condition 2, case I. We can distinguish two options for conservative boundary conditions. If we impose $\d\eps_\cN=0$, then \eqref{Thbc} vanishes for $\d l^\m=\d \s^{\m\n}=0$, which in turns imply $\d\th=0$, and $(2-b)\d k=0$. This gives us Dirichlet boundary conditions for $b=2$, and strengthened Dirichlet conditions including $\d k=0$ for $b\neq 2$. 
If we don't impose $\d\eps_\cN=0$, then necessarily $b=0$ and $c=1$, and we find the conformal boundary conditions \eqref{Ybc} of the York polarization.

Condition 2, case II. If we want stationarity to correspond to a shear and expansion-free surface, which we recall is equivalent to a NEH in vacuum, then we need
\begin{align}\label{Thbc2}
 \big(\pi_\m \d l^\m +(2-b)\d k+(2-c)\d\th \big)\eps_\cN - bk \d\eps_\cN=0.
\end{align}
The most general solution of this equation is 
\be\label{fixeddk}
b=0, \qquad \d l^\m\eqonN0, \qquad \d k\eqonN\f{c-2}2\d\th.
\ee
For $c=2$, we recover the result of \cite{Chandrasekaran:2018aop}: the symplectic potential 
\be\label{CFPth}
\th^{\sscr CFP} = \s^{\m\n} \d \g_{\m\n}  \eps_\cN - \th \d\eps_\cN
\ee
meets all the WZ criteria (and was argued in \cite{Chandrasekaran:2018aop} to be unique under these conditions).
For arbitrary $c$, we find a 1-parameter family of covariant WZ potentials that satisfy the same stationarity condition,
\be
\th^{\sscr c} = \s^{\m\n} \d \g_{\m\n}  \eps_\cN -(c-1) \th \d\eps_\cN.
\ee
This includes the conformal polarization for $c=1$. 
These potentials are associated with a phase space in which inaffinity is allowed to vary, but in a way fully constrained by the expansion via 
\eqref{fixeddk}.
These properties holds also if we further relax $\d l_\m\eqonN0$.
This defines a new phase space that could be interesting to further explore.
The difficulty with this generalization of the CFP result  is that the symmetry vector fields appear to be field-dependent and not universal, as we saw in Section~\ref{SecSym}.

\subsection{Stationarity on flat light-cones}

The notion of stationarity as shear and expansion-free used above is solidly based on physical grounds: shear and expansion-free hypersurfaces capture the idea that no radiation is going through the surface, and include standard stationary examples such as non-expanding horizons and Killing horizons. However, this notion is not exhaustive of stationarity understood as lack of radiation, as there are plenty of null hypersurfaces which possess shear and expansion even in the absence of gravitational waves. Consider for instance a light-cone in flat Minkowski space: its expansion grows, hence the CFP flux \eqref{CFPth} is non-zero, even though there is no actual dynamics taking place. This is an objectable feature, which disconnects charge conservation from absence of radiation.
It motivates the question whether one can find a different potential leading to a vanishing flux on both non-expanding horizons and flat light-cones. 
This is not possible within the framework above, because the CFP symplectic potential is unique under the requests of covariance and stationarity on NEH. What we propose is  to relax the notion of stationarity, from $\th'=0$ to: 

\smallskip

Case III: $I_\xi\th'=0$ for \emph{every} symmetry vector field on the stationary solutions.

\smallskip

Namely, we require vanishing of the Noether flux, as opposed to vanishing of the symplectic flux.
This requirement is weaker than $\th'=0$, therefore the immediate consequence is a lost of uniqueness.\footnote{It is still stronger than the minimal requirement of stationarity for isometries, which is so weak that it is satisfied even by Komar.
}
 While this appears bad at first sight, our point is that it can be compensated by the larger set of solutions that can be included.
For the family of anomaly-free potentials,
\begin{align}\label{weakstationarity}
I_\xi\th^{\sscr c}&= \big[ \s^{\m\n} \d_\xi \g_{\m\n}  + \pi_\m \d_\xi l^\m +2\d_\xi k+(2-c)\d_\xi\th \big]\eps_\cN
-(c-1)\th\d_\xi\eps_\cN \nn\\
&= \big[ \s^{\m\n} \pounds_\xi \g_{\m\n}  + \pi_\m \pounds_\xi l^\m 
+2(\pounds_\xi k-\pounds_l w_\xi)+(2-c)\pounds_\xi\th \big]\eps_\cN
-(c-1)\th\pounds_\xi\eps_\cN.
\end{align}
On a NEH in vacuum $\s_{\m\n}=\th=0$ and $\pounds_\xi\th=0$ for tangent diffeomorphisms, and the flux reduces to
\begin{align}\label{NEHflux}
I_\xi\th^{\sscr c}\stackrel{\sscr{NEH}}= \big(\pi_\m \d_\xi l^\m +2\d_\xi k \big)\eps_\cN
= \big( \pi_\m \pounds_\xi l^\m + 2 (\pounds_\xi k-\pounds_l w_\xi) \big)\eps_\cN.
\end{align}
This vanishes only for $\d_\xi l^\m= \d_\xi k=0$, namely for symmetry vector fields belonging to the CFP group. 
On the other hand, it vanishes for any $c$, hence the weaker stationarity condition leaves an ambiguity in the choice of symplectic potential. 
But this ambiguity is eliminated because we can now extend the set of  solutions that fullfill the stationary property. 
Consider a flat light-cone. Its expansion does not vanish, hence it does not fit into the previous notion of stationarity. Only the shear vanishes, so the flux gives 
\begin{align}\label{flatflux}
I_\xi\th^{\sscr c}\stackrel{\sscr lightcone}=(2-c)\d_\xi\th \eps_\cN -(c-1)\th\d_\xi\eps_\cN = (2-c)\pounds_\xi\th \eps_\cN -(c-1)\th\pounds_\xi\eps_\cN -\th w_\xi \eps_\cN. 
\end{align}
We also recall that super-translations are not allowed on a semi-complete null surface, so the CFP group is here reduced to Weyl transformations and super-Lorentz.
To evaluate the flux, we can exploit the fact that the potential is class-III invariant and make a convenient choice of normal. 
We take affine coordinate $\l$ and 
\be\label{lldl}
l = \l\p_\l, \qquad \Rightarrow \qquad \th=2, \qquad \eps_\cN = \f{\sqrt{-g}}fd^3x=\l d\l\w\os{\eps}_S.
\ee
Then the first term in \eqref{flatflux} vanishes, and so does the anomaly, see \eqref{anaffine}: Only super-translations have non-vanishing anomaly for this choice of $l$, but these are not part of the allowed symmetries.
The only non-zero term is the second one, since $\pounds_\xi\eps_\cN\neq 0$ for any non-trivial $W$ and $Y$. 
The flux on a flat light-cone is thus  $I_\xi\th^{\sscr c} =(1-c)\pounds_\xi\eps_\cN$, and vanishes for arbitrary transformations only for the choice  $c=1$.

We conclude that the flux \eqref{weakstationarity} vanishes for NEH \emph{and also} for a flat light cone if $c=1$.
Stationarity in the weaker sense of case III allows one to solve the problem of a non-vanishing flux on a flat light-cone.\footnote{One may wonder whether this idea can be taken one step further to identify a symplectic potential with vanishing flux on shear-full hypersurfaces in flat spacetime. We don't know if this could be done, but it seems difficulty since even a non-radiative shear can introduce a dynamical evolution of the null hypersurface. The case of a light-cone is special in this sense because even if the area changes, the expansion has a constant representative.} 
This process selects again a unique potential. It is the conformal one instead of the CFP one, and on the CFP phase space reads
\be
\th^{\sscr Conf} =( \s^{\m\n} \d \g_{\m\n} +\d\th) \eps_\cN.
\ee

\subsection{Larger phase spaces}

We have seen that the minimal condition for covariance is $\d l_\m\eqonN 0$. Therefore the larger phase space defined by $l_\m \d l^\m=\d l_\m \eqonN0$ admits a one-parameter family of covariant symplectic potentials. This space contains arbitrary variations of the three tangent components $\d l^\m$ and of $\d k$, and its symmetry group is $\Diff(\cN)\ltimes \R^\cN$. The stationarity condition is violated in both its original WZ definition and the weaker one of case III, see \eqref{Thbc2} and \eqref{NEHflux}.
Adding the restrictions $n_\m\d_\xi l^\m=0$ alone reduces the symmetry group to $\Diff(\cN)$ but breaks covariance, because this condition is not class-I invariant. 
Adding instead $\d l^\m=0$ reduces the group to ${\Diff_l(\cN)}$, 
covariance is preserved for $b=0$ even without $\d l_\m=0$, and stationarity is lost in both versions.

Specifically to the `relaxed' CFP phase space with varying inaffinity considered in the Appendix of their paper, we find that the symmetry vector fields span the group $\bigG$ given in \eqref{bigG}, with $\d l_\m=0$ required to have closure under the Lie bracket, while $n_\m\d_\xi l^\m=0$ is required to make the anomaly `canonical', namely depend on the symmetry parameters and $f$ but not on the extension, see \eqref{BMWanomaly}.
The first condition also guarantees that $\th^{\sscr c}$ with $b=0$ is covariant. 
On the other hand the stationarity condition is violated for both the original WZ definition and the weaker definition of case III. In other words, the vector fields in $\bigG$ which are not in $G^{\sscr CFP}$ have a non-vanishing flux on a non-expanding horizon, given by
$I_\xi\th= 2(\pounds_\xi k-\pounds_l w_\xi -k w_\xi)\eps_\cN$. In affine coordinates this reduces to $f\ddot \t$, making it manifest that it vanishes only for the CFP vectors.

\subsection{Charges} 

In this Section we use the formula \eqref{q'} to write the improved Noether charges for arbitrary variations, arbitrary $\xi$ and any choice of $(b,c)$ in the boundary Lagrangian. We will give some general observations and then comment on the special features that occur when the additional restrictions are added, and specific values of $b$ and $c$ chosen, in parallel with the discussion on the fluxes of the previous Section. In particular we will explain how they relate to the WZ prescription, and when they can be identified as WZ charges.

In the following we take $n$ adapted to the cross sections $S$ of the null boundary on which we are evaluating the charges. This fixes the class-I ambiguity, and it is a choice useful to simplify various expressions.
Namely, 
\be
n=\f1{fg^{\l\Phi}} d\l
\ee
where $\l$ is an arbitrary parameter labelling the cross sections of a space-like foliation of $\l$. 
We can then write the pull-backs as follows. For the Komar charge,
\begin{align}
& q_\xi = -\frac 12 \epsilon_{\m\n\rho\sigma}\nabla^\m\xi^\nu dx^\rho\wedge dx^\sigma \pb{S} 2n_\m l_\n \nabla^{[\m}\xi^{\n]}\epsilon_S. 
\end{align}
For the boundary Lagrangian, 
\begin{align}
& i_\xi\ell^{\sscr (b,c)}= -(bk +c\th)i_\xi\eps_\cN \pb{S}(bk +c\th) \xi\cdot n\,\epsilon_S.
\end{align}
For the corner symplectic potential,
\begin{align}
	I_\xi 	\vth^{\sscr EH} &\pb{S} (n_\m \d_\xi l^\m + n^\m \d_\xi l_\m) \eps_S 
= (n_\m \pounds_\xi l^\m + n^\m \pounds_\xi l_\m + 2 w_\xi )
\eps_S,
\end{align}
where we used \eqref{defw}. 
The Lie derivatives satisfy the following identity,
\be \label{lielietokBY}
n^\m\pounds_\xi l_\m + n_\m \pounds_\xi l^\m = 2n_{\m}l_{\n}\na^{[\m}\xi^{\n]}+2n_\m \xi^\n\na_\n l^\m. 
\ee
The first term on the RHS coincides with the pull-back of the Komar 2-form. The second is a contraction of the Weingarten map, thanks to the restriction of $\xi$ to be tangent.

Adding up according to \eqref{q'}, we get
\begin{align}\label{qbc}
	q_\xi^{\sscr (b,c)} &=	-2 [n^\m \xi^\n (W_{\n \m} - \frac{1}{2}(bk + c\theta)g_{\m \n}) + w_\xi ]\eps_S
 \\\nn &= -[2n^\m \xi^\n (W_{\n \m} - Wg_{\m \n}) + \xi\cdot n((2-b)k+(2-c)\th)+ 2w_\xi] \eps_S 
	\\\nn &= - [ 2\xi^\m( \eta_\m - \th n_\m) + \xi\cdot n((2-b)k+(2-c)\th) + 2w_\xi ] \eps_S.
\end{align}
These are the Noether charges for the full group $\Diff(\cN)\ltimes \R^\cN$, and any polarization in the family \eqref{ellfamily}. 
No restrictions except for $l_\m\d l^\m=0$.
The $c$-term in the boundary Lagrangian is a total derivative and could have been moved to $\vth$. This move leaves the charges invariant, 
because corner shifts in the boundary Lagrangian only matter if the shift is anomalous \cite{Chandrasekaran:2021vyu,Odak:2022ndm}, 
and $\th\eps_\cN$ is not. Using $\ell^{\sscr D}$ versus $\ell^{\sscr D}{}'$ in the case of Dirichlet polarization, or $\ell^{\sscr Conf}$ versus nothing in the case of conformal polarization, is irrelevant.

We can now make the earlier discussion on the need of a physical prescription for the charges  concrete.
First of all, they are in general not class-III invariant, and depend explicitly on the choice of normal representative taken.
Secondly, they depend on the extension $\xiext$ of the symmetry vector fields  through the anomaly $w_\xi$. Therefore if this extension is a free parameter, the charges take value on the group $\Diff(\cN)\ltimes \R^\cN$ and can be given an arbitrary value even if the intrinsic parameters on the hypersurface are kept fixed.
Further problems appear if we look at their flux, which is given by
\begin{align}\label{genflux}
\pbi{dq}{}^{\sscr (b,c)}_\xi &\eqons I_\xi\th^{\sscr (b,c)}-a^{\sscr (b,c)}_\xi \\\nn &=
[\s^{\m\n}\pounds_\xi\g_{\m\n}+\pi_\m\pounds_\xi l^\m +(2-b)\pounds_\xi k-2\pounds_l w_\xi +(2-c)\pounds_\xi \th-2\th w_\xi]\eps_\cN -[bk+(c-1)\th]\pounds_\xi\eps_\cN.
\end{align}
The problem with this flux is the same one that plagues the Komar charges: it can be non-zero flux even on a NEH or in Minkowski space for generic diffeomorphisms tangent to a generic null hypersurface!
This is what we meant in the earlier discussion when we said that a generic version of Noether theorem may be unpractical, and one needs some additional input to reorganize it in a more useful way. To make this more precise, recall that  the dynamical content of the flux-balance laws is the constraint equations, namely for a null hypersurface the Raychaudhuri and Damour equations, as discussed for instance in \cite{Hopfmuller:2018fni}. These can be derived from \eqref{genflux} for 
 $\xi$ tangent to the null geodesics or to the cross-sections, respectively. The point is that for arbitrary $b$ and $c$ the terms in $\dot\th$ and $\dot{\bar\eta}_\m$ will appear scattered on both LHS and RHS, and that without phase space restrictions there will be gauge-dependent terms in both charge and flux that cancel out in the final equation.
Let us know see how these problems are solved using the WZ prescription described above.

The first thing we want to comment on is the explicit appearance of the anomaly $w_\xi$. This responsible for the shift between these charges and a Brown-York-like expression based on the Weingarten map alone, as was found in \cite{Chandrasekaran:2021hxc} for the Dirichlet polarization.\footnote{See also \cite{Brown:1996bw,Jafari:2019bpw} for other Brown-York-like formulas on null boundaries.} One of our initial motivation was to study whether this shift could be removed changing polarization. As we can see from the general expression \eqref{qbc}, this is not the case for the polarizations considered. They only affect the numerical coefficients that would give rise to the trace term $W$, and not the anomaly contribution. But the anomaly term is actually very important: it leads to the area being the charge associated with a constant Weyl rescaling, arguably the most famous gravitational charge for horizons. More in general, it is crucial to guarantee that these charges are perfectly covariant for $b=0$, because it compensates the non-class-III invariance of $\eta_\m$ and $k$.
To see this, we use \eqref{defw} to rewrite the charges as
\be
q_\xi^{\sscr (b, c)} =  -2 \left[\xi\cdot \bar \eta -\xi\cdot n\left(\f c2\th+\f b2 k-\bar k\right) +\xiext-\d_\xi\ln f\right] \eps_S.
\ee
What we have done here is to expand the term $\pounds_l \ln f$ in the anomaly and reabsorb it into the shifts of $k$ and $\eta_\m$ to 
$\bar k$ and  $\bar \eta_\m$.
Recall that the terms in $\bar k$ and $\bar \eta_\m$ are partially invariant under a class-III transformation, namely they change only if the rescaling is induced by a reparametrization of $\Phi$. But the same reparametrization changes also the anomaly, see \eqref{defw}! The two changes perfectly compensate, making these terms fully class-III invariant. The only non-class-III invariant contributions are the terms in $k$ and $\d_\xi l_\m$, and this was to be expected from the results we obtained previously on the covariance of the symplectic potential. We conclude that the charges are class-III invariant for $b=0$ and $\d l_\m=0$. To complete the proof of covariance we need to check also for class-I invariance. This cannot be done for \eqref{qbc} because we wrote it in a fixed choice of rigging vector, but follows from the properties of the symplectic potential, and could be easily checked writing the more general formulas for the pull-backs with a non-adapted rigging.

We have seen the importance of the anomaly in establishing covariance of the charges. This can be stated in other words as follows: writing the charges in terms of geometric quantities such as those that enter the Weingarten map and its decomposition requires the use of non-dynamical fields, and the anomaly contribution is there to remove the non-dynamical dependence. 
One of the consequences of this result is that a Brown-York-like construction for the charges on null hypersurfaces cannot be covariant.

If we now impose $n_\m \d l^\m=0$ as in \cite{Hopfmuller:2018fni}. Working in affine coordinates, so that $\bar k=0$ and we can use the explicit formula \eqref{BMWanomaly}, we obtain
\begin{align}
	q_\xi^{\sscr (b,c)} &=  -2 \left[Y \cdot \bar \eta +\t \left(\f c2\th_\l  +\f b2 k\right) - \dot\t \right] \eps_S.
\end{align}
These are the general charges for the group $\Diff(\cN)$. The above formula gives the illusion that they are covariant for $b=0$ without $\d l_\m\eqonN0$, but this is wrong, covariance is spoiled by the non-class-I invariant restriction that one is making. This is true regardless of the value of $b$ and $c$, hence we conclude that charges for the symmetry group $\Diff(\cN)$ constructed in this way are not covariant. 
While it may be possible to construct other $\Diff(\cN) $ charges, we believe there is an intuitive reason as to why they cannot be covariant. 
First, we have seen that the dependence of the charges on the first-order extension of $\xi$ is crucial to remove the dependence on the embedding, which cannot be eliminated as in the non-null case by choosing the unit-norm normal. So the natural group of charges associated with a null hypersurface is $\Diff(\cN)\ltimes \R^\cN$. Second, there is no canonical way to reduce this group to $\Diff(\cN)$ because of the lack of a projector on a null hypersurface. One can pick a $\Diff(\cN)$ choosing a section of $\Diff(\cN)\ltimes \R^\cN$ via the rigging vector, but there result will depend on the rigging vector, hence non-covariant and anomalous.

To continue the discussion on covariance, we take $b=0$ and consider instead the restrictions $\d l_\m=\d l^\m\eqonN 0$. We then have
\begin{align}\label{qc}
	q_\xi^{\sscr c} &=-2 \left[Y\cdot \bar \eta +\f c2\t\th - \dot\t \right] \eps_S.
\end{align}
This is a one-parameter family of covariant charges for the new group ${\Diff_l(\cN)}$, explicitly class-III invariant. 
In particular: the super-Lorentz charge aspect is the shifted twist $\bar \eta$, and we have charges aspects for the null diffeomorphisms and their first derivative given respectively by the expansion and the area.

If we add the CFP restriction $\d k\eqonN 0$, we can use $\t=T+\l W$ as in the form \eqref{xiCFP} of the symmetry vector fields, and write
\be\label{qcCFP}
q_\xi^{\sscr c} =  -2 \left[Y\cdot \bar \eta +c\, T\th_\l +\left(\f c2\l\th_\l-1\right) W \right] \eps_S,
\ee
where $\th_\l$ denotes the expansion of the affine generator. This is a one-parameter family of covariant charges for the  group $G^{\sscr CFP}$.
We can also remark that the restriction \eqref{ddottau} implies that $\pounds_l \xiext=0$, therefore the partially class-III invariant quantities $\bar k$ and $\bar\eta_\m$ become fully class-III invariant, and accordingly, their anomaly vanishes.
We have the same super-Lorentz charge aspect as before, and the null diffeomorphisms are now split into a super-translation charge aspect given by the expansion, and the Weyl charge aspect given by the area minus the expansion. The latter in particular reduces to the area on a NEH, and we see that this result holds for any $c$.
This is a general property: all charges are both $b$ and $c$ independent on a NEH. Requiring $\cN$ to be a NEH is a huge restriction on the phase space of general relativity, and the fact that a degeneracy in the symplectic flux and charges is introduced should not be surprising.

Now let us talk about stationarity. To understand the need of additional requirements to prescribe the charges, observe that this family of covariant charges contains the original Komar expression! This occurs for $b=c=0$ and $\d l_\m =\d l^\m=0$. Therefore covariance alone can only take us so far, and we still haven't solved the initial problem of little physical meaning. Hence the importance of the stationarity requirement. 
If we require it in the strong sense of Case II, then we must pick  $c=2$ \eqref{qcCFP}, and we recover in this way the CFP charges of \cite{Chandrasekaran:2018aop}.
In the special case of a NEH they are given by
\be
q^{\sscr c}_{\xi} \stackrel{\sscr NEH}{=} -2 [Y\cdot \bar \eta - W ] \eps_S.
\ee
They match those given in \cite{Ashtekar:2021kqj}, and are conserved and independent of the polarization parameter $c$ as stated above. 
A somewhat subprime feature of this charges is that we would like the reference solution in which they vanish to be flat spacetime, but this is not the case, since flat spacetime does not contain NEH of compact cross-sections. What was done in \cite{Chandrasekaran:2018aop} was to evaluate them for a Schwarzschild horizon of mass $M$ and argue that they tend to zero in the limit $M\to 0$. What flat spacetime contains are light-cones, these are shear-free null hypersurfaces of compact cross-sections, but are expanding. On a flat light-cone we have $\bar\eta_\m=0$, and the charges are given by
\be
q^{\sscr c}_{\sscr (W,Y)} \stackrel{{\footnotesize{l.c.}}}=  -2 [(c-1) W ] \eps_S.
\ee
They are not conserved unless $c=1$, in which case they vanish. Therefore if we require stationarity in the weaker sense of Case III, we select $c=1$ and obtain charges that only only are conserved on both NEH and flat light-cones, but also vanish on flat light-cones.
The conformal polarization $c=1$ not only has better stationarity properties, but it also makes it more natural to assess that the reference solution for the charges is flat spacetime.

The choice of charge with $b=0$ and $c=1$ has also an interesting consequence for the relation between its flux-balance law and the Raychaudhuri equation. Recall in fact that if take $\th$ as the charge, its flux $\dot\th$ is not monotonic, even if the null energy conditions are satisfied. When choosing $b=0$ and $c=1$ on the other hand, the Raychaudhuri equation is reorganized so that the charge $2(1-\f12\l\th_\l)\eps_S$ has a monotonic flux if the null energy conditions are satisfied and the hypersurface is future complete \cite{Rignon-Bret:2023fjq}. This fact can be used for notions of dynamical entropy in the context of the generalized second law. Obtaining a monotonic flux from the Raychaudhuri equation is also studied in the recent work \cite{Ciambelli:2023mir}.

Finally, let us cover the relation between these improved Noether charges and the Wald-Zoupas charges. The covariance requirement for the symplectic potential has selected the family with $b=0$. For this family the boundary Lagrangian is anomaly-free, hence we are in case $(a)$ of \cite{Odak:2022ndm}: the covariant improved Noether charges are Wald-Zoupas charges, there is no need for a corner shift. They can however be considered proper Wald-Zoupas charges only for those symmetry groups for which the stationarity condition is satisfied. This means $G^{\sscr CFP}$ and $G^{\sscr AKKL}$, while for the larger groups $\bigG$ and $\Diff(\cN)\ltimes \R^\cN$ the charges are well defined but do not satisfy the stationarity condition neither in the original sense of case II, nor in the weaker sense of case III.

\section{Addenda}

\subsection{Second-order perturbations around flat light-cones}

In the previous Section, we identified covariant charges and fluxes which are conserved on a flat light-cone, and vanish exactly on each cross sections.
This occurs for the special choice of polarization $b=0$ and $c=1$, and for symmetry vector fields belonging to the CFP group.
We now study their evolution when the light-cone is perturbed by gravitational radiation. Since the charges and fluxes are covariant, we can use any normal representative, and we pick the choice \eqref{lldl} with a constant expansion in flat light-cones and vanishing anomaly.
The flux-balance law is 
\be
	dq_\xi = (\s^{\m \n} \pounds_\xi \gamma_{\m \n} + \pounds_\xi \th )\eps_{\cN},
	\label{fluxlcge}
\ee
where we remember that only super-Lorentz $Y$ and Weyl super-translations $W$ are allowed as symmetries. 
For a pure Weyl transformation, 
\be
	dq_{W}= W (2 \s_{\m \n}^2 + \pounds_l\th)\eps_{\cN}.
	\label{fluxlcW}
\ee
We now solve for $\pounds_l \th$ using the Raychaudhuri equation, in a perturbative expansion around a NEH. We assume that the shear is infinitesimal, and write
$\th=2+\th_1+O(\s^4)$. Linearizing the Raychaudhuri equation we find
\be
	\pounds_l \th_1 = -\th_1 - \s_{\m\n}^2,
	\label{Raychlin}
\ee
whose solution is
\be\label{linexpan}
	\th_{1}(\l, x^A) = - \f1\l \int_{\l_0}^\l \s_{\m\n}^2(\l', x^A) d\l'. 
\ee
Here $\l_0$ can be taken as the value of affine parameter after which the perturbation enters the light-cone, with the tip located at $\l_0 = 0$.
Plugging this result in \eqref{fluxlcW} and integrating over a region $\D\cN$ of the null hypersurface, we get
\be
	\D q_{W} = \int_{\Delta \cN}  W \left( \l^3 \s_{\l}^2 + \int_{\l_0}^\l \l'{}^{2} \s_{\l'}^2 d\l'\right) d \l \wedge \os{\eps}_S +O(\s^4),
	\label{fluxlcW2}
\ee
where $\s_\l:=\s/\l$ is the shear of the affinely parametrized normal.
The flux is made of two pieces. The first one, proportional to the shear squared, represents the energy of weak gravitational waves entering the light cone locally. It is a tidal heating term. The second piece is related to the gravitational waves which have entered the light-cone since $\l_0$. Unlike the first term, this terms is not local, and depend on the history of the gravitational waves which entered the outgoing light from $\l = 0$. Hence, even in the absence of local shear, we expect a variation of the charge if some weak gravitational waves have previously entered the outgoing light cone. This is because if it is the case, spacetime is not flat anymore in the local surrounding, as there is some energy localized inside the outgoing light cone, the energy of the gravitational waves which have previously entered. Hence, this second term is a memory effect. Furthermore, the flux of the future pointing diffeomorphisms is positive, and so the charge increases, underlying the fact the gravitational waves carry positive energy. Monotonicity is a key property for flux, and one more reason to appreciate the conformal polarization. Here we have proved it perturbatively, but it  can be extended to all orders under the hypothesis of future completeness of the hypersurface \cite{Rignon-Bret:2023fjq}.

Next, we take a pure super-Lorentz, so $\xi\in TS$. In affine coordinates, $\xi^\m \p_\m = Y^A \p_A$ and the charge density associated to this tangent vector is just given by $q_{Y} = - 2 \bar{\eta}_\m \xi^\m \eps_S = Y^A P_A \eps_S$, with variation
\be
	dq_Y =d(Y^A P_A \eps_S)= - 2 \pounds_l (\xi^\m \bar{\eta}_\m).
	\label{chargevariationangmom}
\ee
 We can now compute the flux of the charge using our flux balance law. We make use of the linearized Raychaudhuri equation \eqref{Raychlin} to express the linearized expansion $\theta_1$ in terms of the shear \eqref{linexpan}. For a tangent diffeomorphism $\xi^\m \p_\m = Y^A \p_A$, we have for small perturbations around the flat light cone
\begin{align}
	I_{\xi} \theta^{\sscr Conf} &= \sigma^{\m \n} \pounds_{\xi} \gamma_{\m \n} \eps_\cN + \pounds_{\xi} \theta \eps_\cN \nn \\
	&= 2 D_\m (\sigma^{\m \n} \xi_\n) \eps_\cN - 2 \xi^\m D_\n \sigma^{\n}_\m \eps_\cN - 
	\left(\frac{1}{\l} \int_0^\l  \sigma^{\m \n} \xi^\r D_\r \sigma_{\m \n} d \l'\right) \eps_\cN \nn + O(\theta_1^2) \nn \\ 
	&= - 2 Y^A D_B \sigma^B_A \eps_\cN + O(\sigma_{\m \n}^2),
	\label{fluxangmom}
\end{align}
where we disregarded the first term in the equation \eqref{fluxangmom} because it was a total divergence which does not give any contribution upon integration on the compact cross sections. Therefore, at leading order, we find that the charge variation is given by the angular derivative of the shear along the cross sections. We notice that the charge variation is proportional to the shear at leading order, not the square of the shear. The coefficient $P_A$ appearing in the charge is the coefficient of the first order expansion of $g_{uA}$ in affine coordinate, and so it has the interpretation of an angular momentum for small perturbations around the flat background. Therefore the charge associated to the tangent diffeomorphsims is modified by the the angular momentum of the weak gravitational waves crossing the outgoing light cone. The equation relating the charge variation \eqref{chargevariationangmom} to the flux is a linearization of the more general Damour equation, derived in the appendix.

These results open an interesting direction which we hope to pursue in future work and link to light-cone thermodynamics \cite{DeLorenzo:2017tgx}.

\subsection{Wald-Zoupas prescription with field-dependent diffeomorphisms}\label{fielddep}

We have seen that if the symmetry vector fields include field-dependent diffeomorphisms, the notions of covariance defined by matching Lie derivatives and by anomaly-freeness can give different answers. The question is then which of the two should be used as condition 1 of the WZ prescription. In our previous paper \cite{Odak:2022ndm} we used the matching of Lie derivatives. Following discussions with Chandrasekaran and Flanagan on the topics of the present paper, we were motivated to provide more details about this choice. In this Section we compare the two options, and show that in the end they are both valid, but with a different definition of charge bracket. We also briefly explain why this difference was in the end not important to understand the application of the formalism to the BMS group at future null infinity studied in \cite{Odak:2022ndm}.

We start from \eqref{Ixiom2}, which is still valid if $\d\xi\neq 0$. 
This formula suggests to take the matching-Lie-derivative options. In fact if we require $(\d_\xi-\pounds_\xi)\th'=0$
\eqref{Ixiom3} is still valid, so we can proceed as before subtracting the preferred flux to obtain an integrable generator.
Furthermore, condition 1 also implies that  the Lagrangian anomaly must be spacetime-exact, specifically that $a'_\xi=ds_\xi$ where now
$\d s_\xi=-q'_{\d\xi}-A'_\xi$  \cite{Odak:2022ndm}. The WZ charges are then defined as in \eqref{WZcharges} with this new $s_\xi$,
and still satisfy the flux-balance laws \eqref{WZfluxes}.
This notion of covariance appears thus naturally when talking about integrability of the charges.\footnote{Looking at \eqref{pFcovariant}, we see that the 
argument for integrability resonates with the `slicing' approach proposed in \cite{Adami:2021nnf}.}

Let us consider now requiring instead $\D_\xi \th'=0$, which as we discussed in Section~\ref{fielddep} is a simpler notion of background-independence. 
Furthermore, it is this property that is satisfied by the standard symplectic potential, $\D_\xi\th^{\sscr EH}=0$, while $I_{\d\xi}\th^{\sscr EH}\neq 0$ in general. 
Imposing the anomaly-free condition, the term $I_{\d\xi} \th'$ appears in the RHS of \eqref{Ixiom2}, and the previous procedure no longer works if this is not zero. 
We can then attempt a charge definition subtracting this new term as well, so that \eqref{Ixiom3} is replaced by
\be\label{Ixiom4}
-\pbi{I_\xi\om} + I_{\d\xi} \th' + d i_\xi \th' = \d I_\xi\th' \eqons \d(dq'_\xi+a'_\xi).
\ee
However we are no longer guaranteed that $a'_\xi$ is spacetime exact.
Therefore this condition alone is not sufficient to define the charges, and must be supplemented by the additional condition that 
\be\label{newcond}
I_{\d\xi} \th'=dX_{\d\xi}
\ee 
for some $X_{\d\xi}$. This additional property suffices to obtain WZ charges when the symmetry vectors are field dependent.
The charges are still given by \eqref{WZcharges}, this time with $\d s_\xi=-q'_{\d\xi}-A'_\xi + X_{\d\xi}$ (up to a closed form), and are 
as before conserved and Hamiltonian generators when $\th'$ vanishes.
Notice that  \eqref{newcond} is guaranteed on-shell if the final boundary Lagrangian is covariant, since we are assuming that  $\d\xi$ is a symmetry vector field.
So one can rephrase the two independent prescriptions $\D_\xi\th'=0$ and \eqref{newcond} also as $a'_\xi=0$ and $A'_\xi=0$ (up to a closed 2-form).
We conclude that even if the notion of anomaly-freeness may appear more natural, it is less economical, in that it is not sufficient per se to guarantee the existence of the WZ charges, and one has to require also \eqref{newcond}. 

A further difference between the two procedure arises if we go beyond the flux-balance properties \eqref{WZfluxes}, and require also that the charges give a representation of the symmetry group. To that end we look again at 
 \eqref{goodcharge}, which for $\d\xi\neq 0$ is replaced by
\be\label{goodcharge2}
\D_\chi q_\xi = \f{\p q_\xi}{\p \xi}\D_\chi\xi = \f{\p q_\xi}{\p \xi}(\d_\chi - \pounds_\chi)\xi = 
q_{\d_\xi\chi} - q_{\llbracket\chi,\xi\rrbracket} = q_{\D_\chi \xi}.
\ee
This is the property satisfied by a Noether charge whose background dependence comes only from $\xi$, such as the Komar 2-form. 
Conversely, the anomaly of a generic Noether charge can be evaluated applying the anomaly operator to the flux formula \eqref{dq2}, which gives \cite{Freidel:2021cjp}
\be
\D_\chi q_\xi \eqons q_{\d_\xi\chi}- q_{\llbracket\chi,\xi\rrbracket} +I_\xi A_\chi +i_\xi a_\chi.
\ee
The result \eqref{goodcharge2} is then associated with a symplectic potential and boundary Lagrangian that are anomaly-free, consistently with the Komar example.
If we take the matching-Lie derivatives requirement for the symplectic potential, we have instead $\D_\chi q_\xi \eqons - q_{\llbracket\chi,\xi\rrbracket}$. 
This different behaviour of the anomaly impacts the representation of the symmetry algebra via the Barnich-Troessaert bracket \cite{Barnich:2011mi}. In the first case the original bracket needs to be modified to subtract the $q_{\d\xi}$ term, along the lines considered for instance in \cite{Freidel:2021cjp}.
Whereas in the latter case the charge algebra is represented by the Barnich-Troessaert bracket without any modification.
This whole discussion is valid up to the addition of closed 2-forms to the charge.

We stress that the relevance of this distinction occurs only when $\d\xi$ is a symmetry vector field.
In the case of future null infinity the symmetry vector fields are field-independent, but it is typical to choose a field-dependent extension which preserves a choice of bulk coordinates. As a result, the Einstein-Hilbert symplectic potential is anomaly-free but not covariant in the sense of matching Lie derivatives,
\be
\D_\xi \th=0, \qquad (\d_\xi-\pounds_\xi)\th = I_{\d\xi}\th \eqons d q_{\d\xi}.
\ee
However, $\d\xi$ is not a symmetry vector field. In this context, we found it clearer to require 
$(\d_\xi-\pounds_\xi)\th'=0$ for the Wald-Zoupas potential defined at $\scri$. This is what we took as definition of covariance in \cite{Odak:2022ndm} and it leads to the correct BMS charges.

\section{Conclusions}

In this paper we have presented a general analysis of the covariant phase space of general relativity on an arbitrary null hypersurface, with arbitrary variations of the metric allowed.
We have computed the fluxes and charges for a two-parameter family of polarizations of the symplectic potential, studied their covariance and conservation properties, and explained their relation to the Wald-Zoupas prescription. 
We pointed out in particular the use of a weaker notion of stationarity that allows to select a unique set of charges that are conserved, and vanish, on a flat light-cone, as opposed to the charges obtained following the strong stationarity condition.
This general analysis should provide useful for many applications of gravitational physics on null hypersurfaces. 

We have reviewed various symmetry groups that arise as variations are restricted and the universal structure is strengthened: $\Diff(\cN)\ltimes \R^\cN$, $\Diff(\cN)$, $\bigG$, $G^{\sscr CFP}$ and $G^{\sscr AKKL}$.
We have highlighted the importance of the anomaly contribution in the expression for the charges. It is necessary to make them independent of the embedding, subtracting off the residual non-class-III invariance of $\bar k$ and $\bar\eta_\m$, and to satisfy the stationarity requirements. This shows one side of the importance of using a  prescription \`a la Wald-Zoupas for the charges, as opposed to the Komar expression or any Brown-York-like expression, which fail to be covariant.
In particular we have shown that covariant charges can be written for all groups except $\Diff(\cN)$. 
The group $\Diff(\cN)\ltimes \R^\cN$ appears naturally because charges need to depend on the first-order extension of the symmetry vector fields in order not depend on the embedding of $\cN$ in spacetime.
The intuitive reason why $\Diff(\cN)$ does not admit covariant charges is that it arises as a section of $\Diff(\cN)\ltimes \R^\cN$, and there is no canonical choice of section because of the lack of a projector on null hypersurfaces. The group $\bigG$ of time-independent super-Lorentz and super-translations of arbitrary time dependence can have interesting applications, for instance we pointed out it allows non-trivial super-translations when $\cN$ has two boundaries, given for instance by intersecting initial and final space-like hypersurfaces. The group $\Diff(\cN)\ltimes \R^\cN$ lacks on the other hand the locality property that charges vanish when the parameters on the 3d boundary vanish, but it is nonetheless argued to be physically relevant in \cite{Chandrasekaran:2023vzb}.

We have used a spacetime description for all quantities, and have found it convenient to describe everything using a NP tetrad. This introduces non-dynamical background quantities that have non-covariant transformation properties in the phase space, but we have shown that independence from the background structure can be easily kept under control. Quantities independent of the choice of NP tetrad are covariant, and can be identified from their invariance under a joint class-I and class-III transformation of the tetrad. Furthermore the extra structures are relevant to the Carroll literature, hence our formalism can be immediately used in that context. 

Understanding covariance as independence from the choice of NP representative has immediate application to clarify the ambiguities of null boundary terms that arise if Dirichlet boundary conditions are imposed in a weak way. We have further discussed why reducing the ambiguity requires working with strengthened Dirichlet boundary conditions, whose meaning is to preserve a choice of affine coordinates on the boundary. Alternatively, the ambiguity can be reduced allowing the inaffinity to change but in a way fixed by the rate of change of the boundary area, namely by the expansion. This choice provides a definition of conformal boundary conditions on null hypersurfaces.

\subsubsection*{Acknowledgements}

We thank Alejandro Perez and Antony Speranza for many discussions on the topics of this paper, and Venkatesa Chandrasekaran and Eanna Flanagan for
discussions and exchanges on our drafts.
We acknowledge support from the John Templeton Foundation via the grant n.$62312$, as part of the
project \emph{The Quantum Information Structure of Spacetime (QISS)}. The opinions expressed in this publication are those of the author and do not
necessarily reflect the views of the John Templeton Foundation.

\appendix

\section{Internal Lorentz transformations}\label{AppLorentz}

The behaviour of all geometric quantities under a class-I transformation \eqref{classI} can be easily computed, or red from \cite{Chandra} using the NP formalism.
To that end, recall that
\be
\s_{\m\n}=-\sigma \bar m_\m\bar m_\n +cc, \qquad \eta_{\m}=-(\a+\bar\b)m_\m+cc. \qquad \th=-2\re(\r),
\qquad k=2\re(\eps).
\ee
The fact that $l$ is hypersurface-orthogonal hence geodesic fixes the two NP coefficients 
$\k=0$ and $\r=-\th/2$. Apart from this, the formulas are general.

\subsection{Class-I}

Under \eqref{classI}, we have:
\begin{align}
& \g_{\m\n}\to \g_{\m\n} + 2\left( a l_{(\m}\bar m_{\n)}+\bar a l_{(\m} m_{\n)}+|a|^2 l_\m l_\n \right), \qquad \eps_\cN\to \eps_\cN, \\
& \s_{\m\n} \to \s_{\m\n}-2(\s \bar a \bar m_{(\m}l_{\n)} +cc) - 2\re(\s \bar a^2)l_\m l_\n
\qquad \s\to\s, \qquad  \th\to\th, \qquad k\to k, \\
& \eta_{\m} \rightarrow\eta_{\m} - [a\bar\s +\bar a\left(k+\r \right)]m_\m+  [\bar a \eta\cdot m -a(a\bar\s+\bar a\left(k+\r\right))]l_\m +cc.
\end{align}
We see that $\eta_{\pbi{\m}}$ is invariant on a non-expanding horizon iff $k=0$.

We now check invariance of the pulled-back standard symplectic potential \eqref{ThN1}. 
We first notice that the corner term $\vth^{\sscr EH}$ is invariant thanks to $m^\m \d l_\m=0$ and the pull-back.
Of the bulk term, the third and fourth are invariant. Plugging the above transformations in the first and second term, and using $l_\m\d l^\m =0$, we obtain
\begin{align}
& (\s^{\m\n}+\f\th2\g^{\m\n})\d\g_{\m\n}\to {\rm idem} +2[(a\bar\s+\bar a \r) m_\m\d l^\m+cc],\\
& 2(\eta_\m +kn_\m)\d l^\m \to {\rm idem} - 2[(a\bar\s +\bar a\r)m_\m\d l^\m +cc],
\end{align}
from which the invariance of $\pbi{\th}^{\sscr EH}$ follows immediately. The result holds also if $\d a\neq 0$.

\subsection{Class-III}

Under \eqref{classIII}, we have 
\begin{align}
& \g_{\m\n}\rightarrow\gamma_{\m\n},\qquad \epsilon_\mathcal N\rightarrow A^{-1}\epsilon_\mathcal N, \qquad
 \s^{\m\n} \to A\, \s^{\m\n}, \qquad \th \to A\,\th, \qquad k  \to  A (k+ \pounds_l \ln A), \\ 
& \eta_\m \rightarrow \eta_\m - \g^\n_\m\na_\n \ln A, \qquad \om_\m \to\om_\m +\p_\m\ln A+l_\m \pounds_n \ln A.
\end{align}

We now check invariance of the pulled-back standard symplectic potential \eqref{ThN1}. 
Using these transformations, we first derive
\be
\vth^{\sscr EH}\to {\rm idem} -2 \d\ln A\, \eps_S.
\ee
It is class-III invariant for a field-independent rescaling, but not for a field-dependent one.
The first of the bulk terms is invariant. The others give
\begin{align}
& -2\om_\m \d l^\m \eps_\cN \to {\rm idem} -2(k\d\ln A+\pounds_{\d l}\ln A) \eps_\cN, \\
& 2\d(\th+k)\eps_\cN \to{\rm idem} +2\Big((\th+k)\d\ln A+\f1A\d(\pounds_l A)\Big)\eps_\cN. 
\end{align}
Adding up, we obtain
\be
\pbi{\th}^{\sscr EH}\to{\rm idem} +2(\pounds_{l}+\th)\d\ln A\,\eps_\cN -2 d(\d\ln A\, \eps_S).
\ee
This is manifestly invariant for a field-independent rescaling. For field-dependent ones invariance follows from the cancellation between the bulk and corner terms thanks to the identity \eqref{anvedi}. 

\subsection{Anomalies and NP representatives}\label{AppAnomaly}

The background structure we use to describe a null hypersurface is a choice of NP tetrad. 
In this Appendix we prove that quantities that are independent of the choice of NP representative, namely invariant under both class I and III transformations, are also anomaly-free. Consider a generic functional $F$ of the dynamical fields $\phi=g_{\m\n}$ and the background fields 
$(\Phi,l_\m,n_\m)$. Anomaly-freeness with respect to $\Phi$ is achieved restricting the diffeomorphisms to be tangent, so we assume to have done that in the following.
The variation of $F$ under a change of tetrad is
\be
\d_{(a,\a)} F = \f{\p F}{\p l} \d_{(a,\a)} l + \f{\p F}{\p n} \d_{(a,\a)} n,
\ee
where
\begin{align}
& \d_{(a,\a)} l=\a l, \qquad \d_{(a,\a)} n= -\a n+\bar a m + {a} \bar{m}, \qquad a\ll 1, \qquad \a\ll 1,
\end{align}
is the infinitesimal version of \eqref{classI} and \eqref{classIII}.
This coincides with the anomalies \eqref{defw} and \eqref{anorig} for $\a=-w_\xi$ and $a =m\cdot Z$. Taking this special values,
\be
\d_{(a,\a)} F = \f{\p F}{\p l} \D_{\xi} l + \f{\p F}{\p n} \D_{\xi} n = \D_\xi F.
\ee
Therefore the vanishing of the LHS implies that $F$ is anomaly-free.

\section{Alternative polarizations}\label{AppPolar}

Different choices of $\th'$ can be obtained  integrating by parts in field space writing $p\d q=\d(pq)-q\d p$ for one or more canonical pairs, or by integrating by parts on the hypersurface and thus moving terms in and out of $\vth$. 
Not all such manipulations are useful when looking for admissible boundary conditions, because they may lead to a symplectic potential which is not in diagonal form, or whose $\d q$ are not independent. 
In the main text we restricted attention to changes of polarization in the spin-0 sector only. In this Appendix we present two more changes that affect the boundary Lagrangian.
We will  not consider integrations by part in spacetime, namely the corner term $\vth^{\sscr EH}$ is always the same. We can then start from \eqref{ThD}. 

Changing polarization in the spin-1 sector can be done using
\be
\pi_\m\d l^\m\,\eps_\cN = \d((\th-2k)\eps_\cN) -l^\m\d(\pi_\m\eps_\cN).
\ee
This manipulation changes the boundary Lagrangian. It remains in the family \eqref{ellfamily}, but with different numerical coefficients. 
As mentioned in the main text, the momentum $\pi_\m$ is determined in terms of the shear of the null hypersurface by the Einstein's equations. Therefore it cannot be specified independently from the shear, hence boundary conditions based on this polarization would be consistent only if the boundary equations of motions are satisfied.

For the second change, we observe that the spin-2 and spin-0 sectors can be written in terms of a single tensor, so that
\be\label{ThNPi}
\pbi{\th}^{\sscr D} =[\Pi^{\m\n} \d \g_{\m\n}  + 2 (\eta_{\m}-\th n_\m)\d l^\m \big]\eps_\cN,
\ee
where
\be
\Pi^{\m\n} = B^{\m\n} - (\th+k)\g^{\m\n} = \s^{\m\n} -\f12\left(\th + 2k \right)\g^{\m\n}.
\ee
Notice the change in the spin-1 momentum, due to \eqref{trecarte}. 
The expression \eqref{ThNPi} for the symplectic potential appears in \cite{Chandrasekaran:2021hxc}, where the restriction $\d l_\m\eqonN 0$ is used, which affects the corner term. The dictionary to compare the expressions is that they use hypersurface indices only, and our $(B,\eta)$ are there called $(K,-\varpi)$.
It is also the expression used in \cite{Ciambelli:2023mir}, where it is referred to as `canonical'.
Written in this form, one can consider a change in polarization of the first term, which is given by
\be
\Pi^{\m\n} \d \g_{\m\n} \eps_\cN = -\g_{\m\n}\d\tl\Pi^{\m\n} d^3x - \d\big(\left(\th + 2k \right)\eps_\cN\big).
\ee
One can also combine this with a change in the spin-1 sector via
\be\label{s1}
(\eta_\m -\th n_\m)\d l^\m \eps_\cN = \d (\th\eps_\cN) - l^\m\d\big((\eta_\m -\th n_\m)\eps_\cN\big),
\ee
and get another element of the same family.
If we now restrict the variations to $\d l^\m=0$ we have
\be
\pbi{\th}^{\sscr EH} = -\g_{\m\n}\d\tl\Pi^{\m\n} d^3x -\d(\th\eps_\cN)+ d\vartheta^{\sscr EH}.
\ee
This is reminiscent of the Neumann form of the symplectic potential in the non-null case, however $\Pi$ misses the $\eta$ part of the extrinsic geometry. 
One could try to resolve this rewriting in terms of $W_{\m\n}$, which gives
\be
\pbi{\th}^{\sscr EH} = [\g^{\m\n}\d W_{\m\n} +\d W + (2\eta_\m+kn_\m)\d l^\m+ kn^\m\d l_\m]\eps_\cN+ d\vartheta^{\sscr EH}.
\ee
So even if $W_{\m\n}$ contains the eta term missing in $\Pi$, it drops out because 
$\eta_\m l_\n \d\g^{\m\n}=0$.

\section{Closure of Lie brackets} \label{Lieclosure}

In this Appendix we study the conditions under which symmetry vector fields are closed under the spacetime Lie bracket.
We consider first the vector fields \eqref{xiMyers}.
First of all we check that they close as intrinsic vectors on $\cN$. Namely we define
\be
\hat{\xi}^\m :=\tau(\l, x^B) \p_\l + Y^A(x^B) \p_A.
\ee
Then we have
\be\label{comm1}
[\hat\xi_1, \hat\xi_{2}] = \hat\xi_{12}, \qquad
\t_{12}:=\t_1\dot\t_2 +Y_1[\t_2] -(1\leftrightarrow 2), \qquad Y_{12}=[Y_1,Y_2]_S.
\ee
The algebra closes and has a semi-direct product structure with $\Diff(S)$ acting on the space $\Diff(\R)^S$ of $\Diff(\R)$-valued functions as scalars.

Next, recall that the condition $n_\m\d_\xi l^\m\eqonN 0$ partially constraints the extension $\xi$ of the intrinsic vectors $\bar\xi$ off of $\cN$, specifically the component $\xi^\Phi=\Phi\xiext=\Phi(\d_\xi \ln{f}-\dot\t)$. The fact that the extension is not arbitrary makes closure of the $\xi$'s under the spacetime Lie bracket not automatic. 
The non-trivial component to check is
\begin{align}
	[\xi_1, \xi_2]^\Phi  &= \Phi((\pounds_{\xi_1} \d_{\xi_2} - \pounds_{\xi_2} \d_{\xi_1}) \ln{f} -\dot \tau_{[\xi_1, \xi_2]} )+O(\Phi^2)\nn \\
	&= \Phi(\d_{[\xi_1,\xi_2]}\ln f -\dot \tau_{[\xi_1, \xi_2]}+(\pounds_{\xi_1} \d_{\xi_2} - \pounds_{\xi_2} \d_{\xi_1}-\d_{[\xi_1, \xi_2]}) \ln{f}   )+O(\Phi^2),
\end{align}
so closure occurs for $\d_\xi f=0$. If $\d_\xi f\neq 0$, it is still possible to obtain closure, but only if $f$ and $\xi$ satisfy 
\begin{align}
(\pounds_{\xi_1} \d_{\xi_2} - \pounds_{\xi_2} \d_{\xi_1}-\d_{[\xi_1, \xi_2]}) \ln{f} &
=(\pounds_{\xi_1} \D_{\xi_2} - \pounds_{\xi_2} \D_{\xi_1}- \D_{[\xi_1,\xi_2]}) \ln{f} \nn\\ &=
\pounds_{\xi_1} \left(\frac{\p \ln{f}}{\p g_{\m \n}} \right)\pounds_{\xi_2} g_{\m \n} - \pounds_{\xi_2}\left( \frac{\p \ln{f}}{\p g_{\m \n}} \right)\pounds_{\xi_1} g_{\m \n}.
\end{align}
This is a non-trivial equation that cannot be satisfied for general $f$ without restricting the diffeomorphisms. In particular we notice that it is satisfied if $\D_\xi f=0$.

Adding the condition $\d_\xi l_\m\eqonN 0$ eliminates the extra term, and the algebra closes. Notice that it closes for $\t$ an arbitrary function on $\cN$, namely for the group $\bigG$ associated with the relaxed CFP phase space with $\d_\xi k$ arbitrary, as well as for $\Diff(\cN)$ associated with the further relaxation of $\d_\xi l^\m$ to  $n_\m \d_\xi l^\m=0$ only. 
This result on the closure of the algebra may appear in tension with \cite{Ciambelli:2021vnn}, where it was proved that the largest \emph{corner} subalgebra that closes at first order under the Lie bracket includes at most a linear dependence in time, as opposed to the arbitrary time dependence of $\t(\l,x^A)$ here. The difference is that the corner subalgebra includes two independent sets of super-translations, corresponding to the two null times that tick off the corner. Restricting the diffeomorphisms to be tangent to $\cN$ eliminates from the algebra those elements that prevent an arbitrary time dependence.

As for the group $\Diff(\cN)\ltimes\R^\cN$ mentioned in the paragraph after \eqref{xigen}, the abelian character of the extra factor follows from the fact that vector fields labelled only by the first-order transversal extension $\xiext$ commute under the spacetime Lie bracket.

\section{Derivation of Damour's equation}

In this Appendix we give for convenience the derivation of Damour's equation in our notation. This equation is the tangential constraint on a null-hypersurface, and its relevance is that it the dynamical content of the flux-balance law for $\bar\eta_\m$.
We fix a cross-section $S$ of $\cN$ at the help of a rigging vector $n_\m$. We then consider a vector $v\in TS$, and contract the Einstein's equations to look at the two tangential constraints:
\begin{align}
	G_{\m \n} l^\m v^\n &= R_{\m \n} l^\m v^\n 
	= v^\n (\nabla_\rho \nabla_\n - \nabla_\n \nabla_\rho)l^\rho.
\end{align}
For the second term we can use \eqref{Wtrace}, and we choose $l$ such that $n^\m \p_\m l^2 = 0$. For the first term, we expand
\begin{align}
	\nabla_\n l^\rho = (\sigma_\n^\rho + \f12 \theta \g_\n^\rho) - l^\rho \eta_\n - k n_\n l^\rho - A^\rho l_\n + B^\rho l_\n,
\end{align}
where $A^\rho = \g_\sigma^\rho n^\m \nabla_\m l^\sigma$ and $B^\rho = l^\rho n_\sigma n^\m \nabla_\m l^\sigma$. Then 
\begin{align}
v^\n \nabla_\rho \nabla_\n l^\rho 
 	&=  v^\n \nabla_\rho  \sigma_\n^\rho + \f12 v^\rho \nabla_\n (\theta \g_\rho^\n)  - (\theta + k) v^\n \eta_\n - v^\n l^\rho \nabla_\rho \eta_\n  - k v^\n l^\rho \nabla_\rho n_\n - A^\rho v^\n (\sigma_{\n \rho} + \f12 \g_{\n \rho} \theta),
\label{derlbynu}
\end{align}
where we used $B^\rho v^\n \nabla_\rho l_\n =  n_\sigma n^\m \nabla_\m l^\sigma (l^\rho v^\n \nabla_\rho l_\n) =  0$. 
Next, we compute
\begin{align}
	 v^\n \nabla_\rho  \sigma_\n^\rho 
	&=  v^\n D_\rho \sigma_\n^\rho + v^\n l^\sigma \sigma^\rho_\n \nabla_\sigma n_\rho + A^\rho v^\n \sigma_{\n \rho }
	\label{derivshear}
\end{align}
where we used a resolution of the identity in $\r$, and $D_\m v^\n:=\g_{\m}^{\r} \na_\r v^\n$ is the 2d covariant derivative, 
and
\begin{align}
	\f12 v^\n \nabla_\rho (\theta \g^\rho_\n)  = \f12 v^\rho \p_\rho \theta + \f12 \theta v^\n \pounds_n l_\n + \f12 \theta v^\n \pounds_l n_\n
	\label{derivexpan}
\end{align}
Furthermore, we write that $-v^\n l^\rho \nabla_\rho \eta_\n = - v^\n \pounds_l \eta_\n + v^\n \eta_\rho \nabla_\n l^\rho =  - v^\n \pounds_l \eta_\n + v^\n \eta_\rho(\sigma^\rho_\n + \f12 \theta \g^\rho_\n)$. If we consider the second term of this equality, it can be divided into two subterms, one proportional to the shear which combines with the term $ v^\n l^\sigma \sigma^\rho_\n \nabla_\sigma n_\rho$ of \eqref{derivshear} and a second term proportional to the shear which combines with the last term of \eqref{derlbynu} to get $- \f12 \theta v^\n \pounds_n l_\n$, and therefore cancels the last term of \eqref{derivexpan}. Thus, we have 
\begin{align}
	v^\n \nabla_\rho \nabla_\n l^\rho = v^\n D_\rho \sigma_\n^\rho + v^\n \sigma^\rho_\n \pounds_l n_\rho + \f12 v^\n \p_\n \theta + \f12 \theta v^\n \pounds_l n_\n - \theta v^\n \eta_\n - v^\n \pounds_l \eta_\n - k v^\n \pounds_l n_\n 
\end{align}
Furthermore, we have that $\pounds_l (v^\m n_\m) = 0$, so if we assume that $n_\m\pounds_l v^\m = 0$, we have that $v^\m \pounds_l n_\m=0$ for any  $v\in TS$, and we get 
\be
	v^\n \nabla_\rho \nabla_\n l^\rho = v^\n D_\rho \sigma_\n^\rho + \f12 v^\n \p_\n \theta  - \theta v^\n \eta_\n - v^\n \pounds_l \eta_\n
\ee
and so by substructing the piece $-v^\n \nabla_\n \nabla_\rho v^\rho$ we arrive at the Damour equation
\be
	T_{\m \n} l^\m v^\n \eps_\cN =  v^\n D_\rho \sigma_\n^\rho \eps_\cN - v^\rho \p_\rho (k + \f12 \theta) \eps_\cN - v^\m \pounds_l (\eta_\m \eps_\cN).
\ee


\providecommand{\href}[2]{#2}\begingroup\raggedright\endgroup


\end{document}